%% file: main.tex
\documentclass[twocolumn, secnumarabic,  floatfix, aps, prl, nofootinbib, nobalancelastpage, longbibliography]{revtex4-2}

\usepackage{xr,ulem}
\usepackage{amsmath, graphicx, array, amssymb}
\usepackage[usenames,dvipsnames]{color}
\usepackage[colorlinks=true,allcolors=Black]{hyperref}

\newcommand{\ETF}{\widetilde E}
\newcommand{\NNC}{\widetilde N}
\newcommand{\EDC}{\widetilde E_{\textrm{d}}}
\newcommand{\ENC}{\widetilde E_{\textrm{nc}}}

\input macros

\begin{document}

\title{Approximate normalizations for approximate density functionals}

\author{Adam Clay$^1$}
\author{Kiril Datchev$^1$}
\author{Wenlan Miao$^2$}
\author{Adam Wasserman$^{2,3}$}
\author{Kimberly J. Daas$^4$}
\author{Kieron Burke$^{4,5}$}
\affiliation{$^1$Department of Mathematics, Purdue University, West Lafayette, IN 47907, USA}

\affiliation{$^2$Department of Physics and Astronomy, Purdue University, West Lafayette, IN 47907, USA}

\affiliation{$^3$Department of Chemistry, Purdue University, West Lafayette, IN 47907, USA} 

\affiliation{$^4$Department of Chemistry, University of California, Irvine, CA 92697, USA}

\affiliation{$^5$Department of Physics and Astronomy, University of California, Irvine, CA 92697, USA}

\begin{abstract}
It seems self-evident that a density functional calculation should
be normalized to the number of electrons in the system.  We present
multiple examples where the accuracy of the approximate energy is
improved (sometimes greatly) by violating this basic principle.  In
one dimension, we explicitly derive the appropriate correction to the
normalization.  Beyond one dimension, Weyl asymptotics for energy levels yield these
corrections for any cavity.  We include examples with Coulomb potentials and the exchange energy of atoms to illustrate relevance to
realistic calculations.
\end{abstract}
\maketitle


It is a truth universally acknowledged, that any density functional
calculation should yield a density that integrates to the number
of electrons in the system. No matter how little is known about the functionals involved,
this truth is so well fixed in the minds of practitioners that the normalization step passes almost unnoticed \cite{austen}.  

Sophisticated approximations to the exchange-correlation 
functional of Kohn--Sham DFT \cite{HohKoh-PR-64,KohSha-PR-65}, 
together with improved algorithms and powerful computers, allow for efficient and accurate  calculations on systems with thousands of atoms~\cite{DelSchWenMesHutVan-CPC-15}. With advances in quantum embedding methods \cite{WasPav-IJQC-20} and orbital-free DFT \cite{MiLuoTriPav-CR-23}, even millions of atoms can be treated \cite{DawDegSteNajRatGen-WIR-22}, with applications ranging from molecular biology~\cite{ColHin-JPCM-16} to drug design \cite{AdeChiAdeSadHamRay-Pha-22} and materials engineering \cite{HorDwaPer-NCS-21}. 
But in the sixty years since the foundational papers, it has never been questioned that, even when minimizing an approximate energy functional, the best normalization constraint is to require that $\int  d {\bf r} \, n({\bf r}) =N$, i.e. the density integrates to the  number of electrons in the system.
 However, motivated by recent advances in the semiclassical study of DFT \cite{BerBur-JPA-20, OkuBur-arxiv-21}, we show here
 that an approximate normalization $\int   d {\bf r} \, n({\bf r}) = \NNC = N + \Delta N$, derived from asymptotic considerations, yields much better energetics than the usual norm.    Remarkably,  changing just this normalization and nothing else often yields better results
 than the same functional evaluated on the exact density (the basic idea behind density-corrected DFT improvements~\cite{SimSonVucBur-JACS-22,VucSonKozSimBur-JCTC-19,SonVucSimBur-JCTC-22,KimSimBur-PRL-13}).
This also represents the generalization of many semiclassical DFT results beyond one dimension. \cite{OkuBur-arxiv-21,OkuCanBur-PRB-24,ArgRedCanBur-PRL-22,EllPitGroBur-PRA-15,ConFabLarDel-PRL-11}
 
\begin{figure}[h]

\includegraphics[width=72mm]{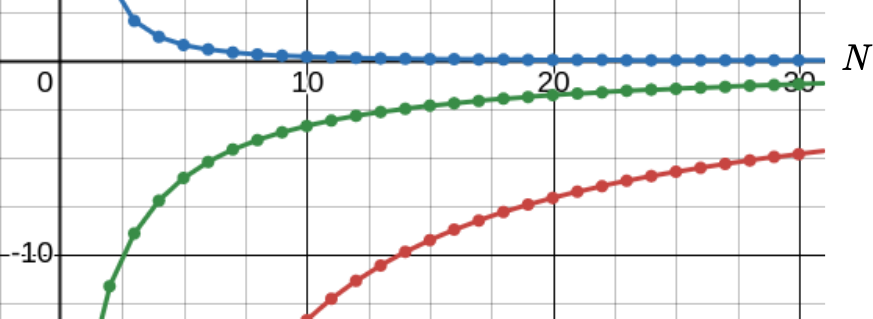} 

\vspace{3mm}

\includegraphics[width=72mm]{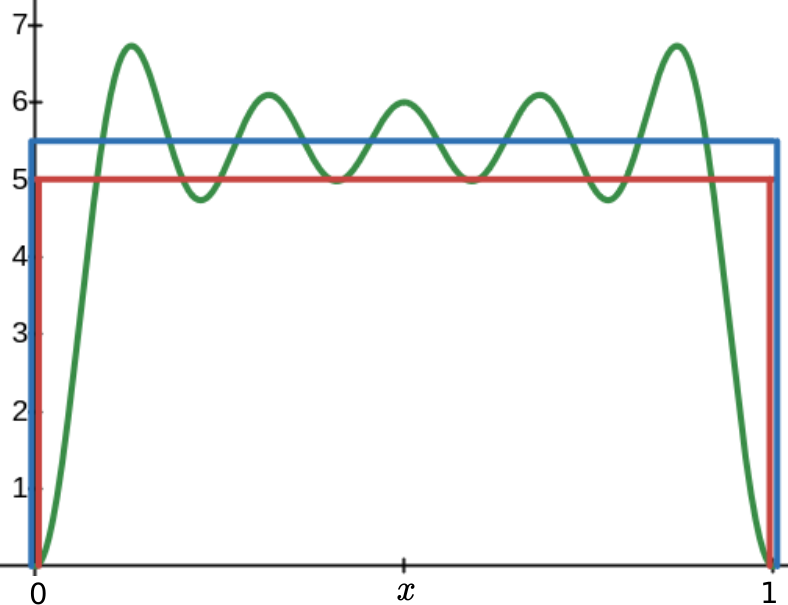}  

\vspace{-1mm}

\caption{
\textit{Bottom}: Red is the  TF density, green the exact density, and blue the normalization-corrected density for $N=~5$. \textit{Top}: Percent errors for the TF energy evaluated on each density (See Table S1 in \cite{SM}).}
 \label{f:pboxintro}
\end{figure}

We start with a simple example: $N$ noninteracting, spinless electrons in one dimension~\cite{RibLeeCanEllBur-PRL-15}.
The local density approximation for such problems is the Thomas-Fermi approximation for the kinetic energy~\cite{Tho-MPCPS-27,Fer-MPCPS-27}, here
$\pi^2 \int dx\, n(x)^3/6$ (using Hartree atomic units).  For the simplest case, the infamous particle(s)-in-a-box,
with  the potential inside the box $v(x)=0$ and box length $L=1$,
Fig.~\ref{f:pboxintro} compares three approximations for the energy $E(N)$. 
The first is the standard DFT 
treatment, where $n(x)$ is found by minimizing the functional, so $n(x) = N$. This energy is denoted $ \ETF(N)$ (tilde signifies an approximation),
and shown in red in the top panel, with its density in the bottom.
 The green is the TF functional evaluated on the exact density, with its energy denoted by $\EDC(N)$ \cite{EllLeeCanBur-PRL-08}.
The blue  is normalization-corrected TF (ncTF),

\vspace{-6mm}

\begin{equation}
 \ENC(N) =  \ETF(\NNC), \quad \NNC=N+ \Delta N,
\end{equation}

 where $\NNC$ denotes an approximate normalization and where $\Delta N=1/2$, outperforming $\ETF(N)$  and  $\EDC(N)$ in all cases. Energy expressions can be found in Section~S1 of the Supplemental Material~\cite{SM}. The normalization correction makes the approximate density closer to the true density  far from the walls, an idea that dates back to Scott \cite{Sco-PMJC-52}.

 In this first example, the correction $\Delta N$ can be derived from the explicit formula for the density $n(x) = \NNC - r(x)$, $r(x)= \sin(2 \NNC \pi x) / 2\sin(\pi x) $ is a lower order term on average, away from the walls. Ignoring the density oscillations yields a better energy than accounting for them and has the advantage of staying within the family of densities belonging to the TF functional.
In mathematical terms, the TF densities of differing   $\NNC$ form a \textit{foliation} of the graph space \cite{CanCon-BOOK-99}.  
For a box or cavity, TF densities are constants and our correction yields the constant which best approximates the bulk, at the expense of the edge. Our approximate constraint beats the exact constraint and even beats using the exact density.

\begin{figure}[h]
 \includegraphics[width=8.5cm]{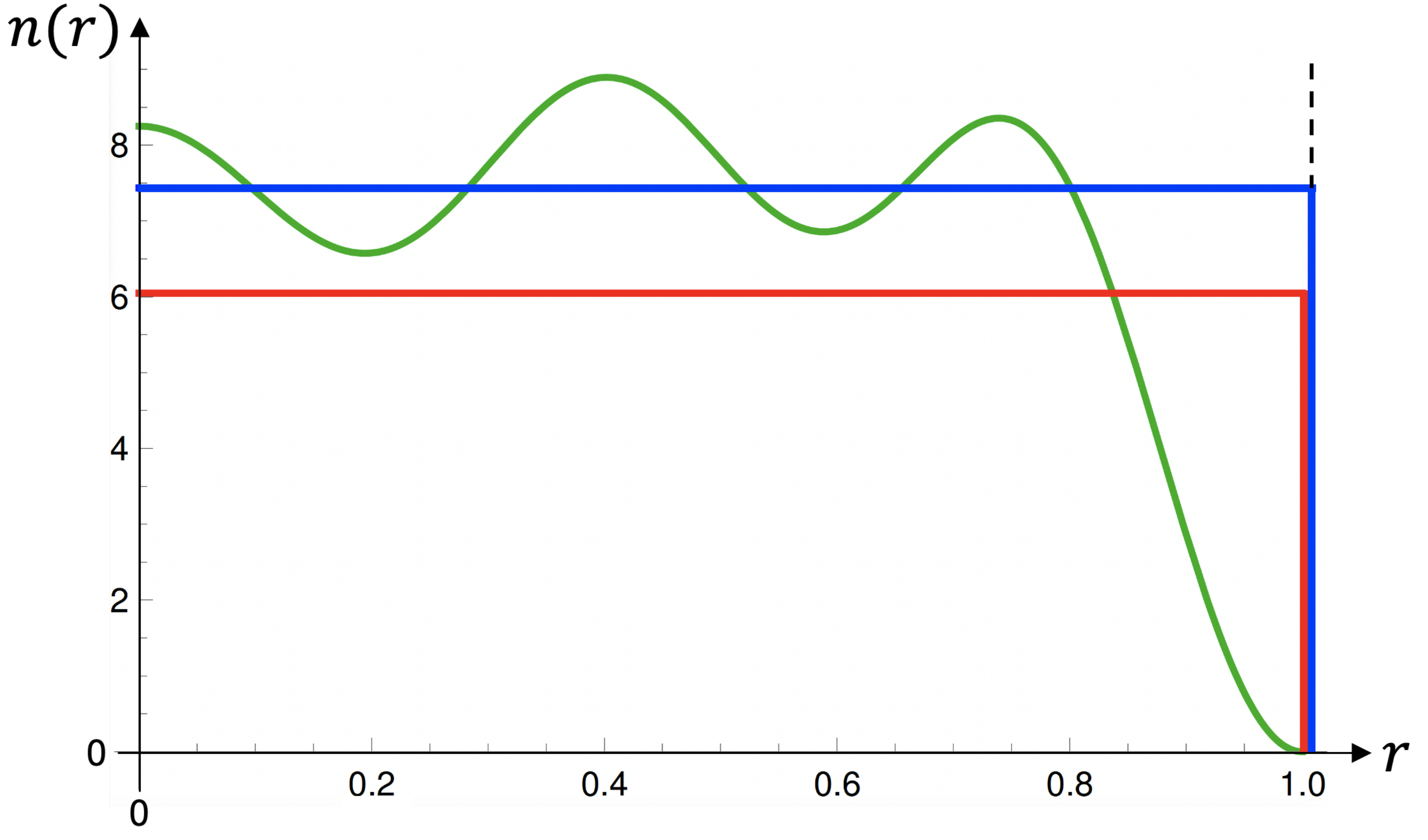}
\caption{ Densities for 19 electrons (11 filled shells) in a circular cavity of radius~$1$. Green is the exact density, red is TF, and blue is~ncTF (Sec. S7 of \cite{SM}, which includes \cite{watson}). An analogous phenomenon, also derived from Weyl asymptotics, is observed in \cite[Fig.~2]{CorsoFriesecke}.
} \label{f:circulardensities}
\end{figure}

This paper presents a proof of principle for our method, focusing on the non-interacting TF kinetic energy functional \cite{Tho-MPCPS-27, Fer-MPCPS-27}, which was first mathematically analyzed in \cite{LieSim-PRL-73, Lie-RMP-81}. We first derive and generalize the 1D example above, using WKB theory~\cite{BenOrs-BOOK-99}. We next explain how Weyl asymptotics for energy levels in cavities can be used to derive values of $\Delta N$ in higher dimensions, with various examples; for instance, Fig.~\ref{f:circulardensities} is the analog of the bottom panel of Fig.~\ref{f:pboxintro} for a 2D circular cavity (see also \cite{CorsoFriesecke}).  We also give several examples for specific
simple potentials, ending with interacting systems and the exchange energy, to connect to realistic DFT calculations.

{\em Proof in one dimension:} For the Hamiltonian $-(1/2) d^2/dx^2 + v(x)$, the TF and WKB approximations are essentially equivalent \cite{MarPla-MPS-56}. From WKB \cite[(1.308)]{Fri-GTP-17}, under reasonable conditions on $v(x)$ \cite{Tit-QJM-54}, we have an implicit formula for individual energy levels
\begin{equation}\label{e:wkb}
 \int_{-\infty}^\infty \frac {dx}\pi p(\mathcal E,x) = \lambda(j) := j-\nu + r(j),
\end{equation}
where $p(\mathcal E,x) = \sqrt{2(\mathcal E-v(x))_+}$ is the classical momentum, $j$ is a positive integer,  $\nu$ is the Maslov index~\cite{MasFed-BOOK-81} ($0$ if there are only hard walls, and increases by $1/4$ for each classical turning point), and $r(j)$ is a remainder which vanishes at least as fast as $j^{-1}$. The exact levels are
\[\mathcal E_j=\mathcal E(\lambda(j))=\mathcal E(j-\nu + r(j)),\]
where $\mathcal E(\lambda)$ solves Eq.~\eqref{e:wkb}.
Summing over $N$ levels yields, for the dominant behavior as $N \to \infty$,
\begin{equation}\label{e:enint}
E(N) =\sum_{j=1}^N\mathcal E_j \sim \int_0^N d \lambda \,\mathcal E(\lambda).
\end{equation}
By changing variables $\lambda \to \mathcal E$, differentiating \eqref{e:wkb}, and swapping the order of integration, \cite{MarPla-MPS-56} showed that $
 \int_0^N d \lambda \,\mathcal E(\lambda) = \ETF(N)$,  precisely  \cite{CanLeeEllBur-PRB-10}.

 An exact version of \eqref{e:enint}, similar to \cite{Bur-FD-20}, is
\begin{equation}\label{e:1dexact}
 E(N) = \int_{1/2}^{N+1/2} \!\! d\alpha \, \Big(\mathcal E(\lambda(\alpha)) + s(\alpha) \frac d {d\alpha}\mathcal E(\lambda(\alpha))\Big),
\end{equation}
where  $s(\lambda) = \lambda - \lfloor \lambda \rfloor - 1/2$ is a saw-tooth  function, and we have extended $\lambda(\alpha)$ from \eqref{e:wkb} smoothly to noninteger $\alpha$. To check \eqref{e:1dexact}, integrate by parts in the second term, and simplify.
Assuming certain derivatives of $\mathcal E(\lambda)$ and $r(\alpha)$ are well-behaved, from \eqref{e:1dexact} the leading correction to \eqref{e:enint} is
\begin{equation}\label{e:1dderiv}
E(N) \sim \int_{0}^{N+1/2 - \nu } d\lambda\, \mathcal E(\lambda) = \ETF( N+1/2 - \nu ).
\end{equation}
This yields  $\Delta N = 1/2 - \nu$, so $\Delta N = 1/2$ in the two-wall case of the introduction (first row of Table \ref{t:ab}) and $\Delta N = 1/4$ in the one-wall case (second and third rows of Table \ref{t:ab}).   This completes the derivation of $\Delta N$ for any potential in 1D, with any non-periodic boundary condition. 
For a harmonic oscillator, $\mathcal E(\lambda) = \omega \lambda$, $\nu = 1/2$, 
for a particle in a box, $\mathcal E(\lambda) = \pi^2  \lambda^2/2 L^2$, $\nu=0$, and  
for a linear half well $\mathcal E(\lambda) = (3 \pi \lambda/2)^{2/3}$, $\nu = 1/4$.

\begin{figure}[ht]
\includegraphics[width=8cm]{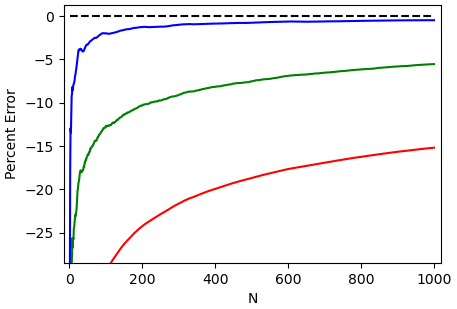}



\caption{Red is energy-minimized TF, green is TF on the exact density, and blue is  normalization-corrected TF, for a  $1 \times \sqrt2 \times \pi $ box.}
\label{f:2x1}
\end{figure}

{\em Weyl asymptotics:} 
In higher dimensions, Weyl asymptotics~\cite{Weyl-11,Wey-crll-13} provide precise information about energy levels for many Hamiltonians, including general classes of potentials and cavities~\cite{ivrii100}. To derive a good $\Delta N$ from such asymptotics is simple.  For
 $N$ noninteracting, spinless electrons in a $d$-dimensional cavity\kim{,} Weyl asymptotics state that
\begin{equation}\label{e:weyl}
 E(N) = C_1 N^{1+ 2/ d} + C_2 N^{1 + 1/ d} + \cdots,
\end{equation}
where $C_1$ and $C_2$ depend on the geometry of the cavity in a simple, explicit way 
\cite[Eq. (6)]{HarProStu-IMRN-19}; see also \cite{CorsoFriesecke}. Related asymptotics hold for other Hamiltonians \cite{AreNitPetSte-MAEIC-09, ivrii100}. The TF approximation yields precisely the first term only, 
and we choose $\NNC$ to recover the second as $N \to \infty$.
Thus, 
$ \ETF(N)$ corresponds to a one-term Weyl asymptotic and $\ENC(N) = \ETF(\NNC)$ to a two-term Weyl asymptotic. 
Table~\ref{t:ex} and Fig.~\ref{f:2x1} show the accuracy of  $\ENC(N)$ for a 3D box.

\begin{table}[h!]
{\setlength{\extrarowheight}{3pt}
    \begin{tabular}{|c|c|c|c|c|}
    \hline
   $N$ &    $E(N)$ &    $\ETF(N)$  & $\EDC(N)$ &  $\ENC(N) = \ETF(\NNC)$ \\
  \hline
       $1$ & $7.90$  &  $  1.69 \ (-79\%)$ & $3.61 \ (-54\%)$ &  $ 5.38 \ (-32\%)$  \\ 
    \hline  
       $10$ & $161$  &  $  78.3 \ (-51\%)$ & $120 \ (-26\%)$ &  $ 148 \ (-8\%)$ \\ 
    \hline  
       $100$ & $5141$  &  $  3633 \ (-29\%)$ & $4490 \ (-13\%)$ &  $ 5039 \ (-2\%)$ \\ 
    \hline  
       $1000$ & $198838$  &  $ 168647 \ (-15\%)$ & $187810 \ (-6\%)$ &  $ 197873 \ (-0.5\%)$  \\ 
    \hline  
    \end{tabular}
}
\caption{Exact and approximate energies for a 3D box with incommensurate edges $1 \times \sqrt2 \times \pi $. See Fig.~3 and S3 of~\cite{SM}.}
\label{t:ex}
\end{table}

Weyl asymptotics were conjectured over a century ago in \cite{Wey-crll-13} and proved in  \cite{DuiGui-IM-75, Ivr-FAA-80}.    They apply to problems in acoustics and black-body radiation, and are important throughout mathematics and physics~\cite{AreNitPetSte-MAEIC-09}. The relevant quantities are averaged versions of the traditional asymptotics, which have been recently proven for very general cavities  \cite{FraGei-MRQP-11,FraGei-BMS-12, FraLar-AM-19, FraLar-Arx-24}.

\begin{figure}[ht]
\includegraphics[width=8cm]{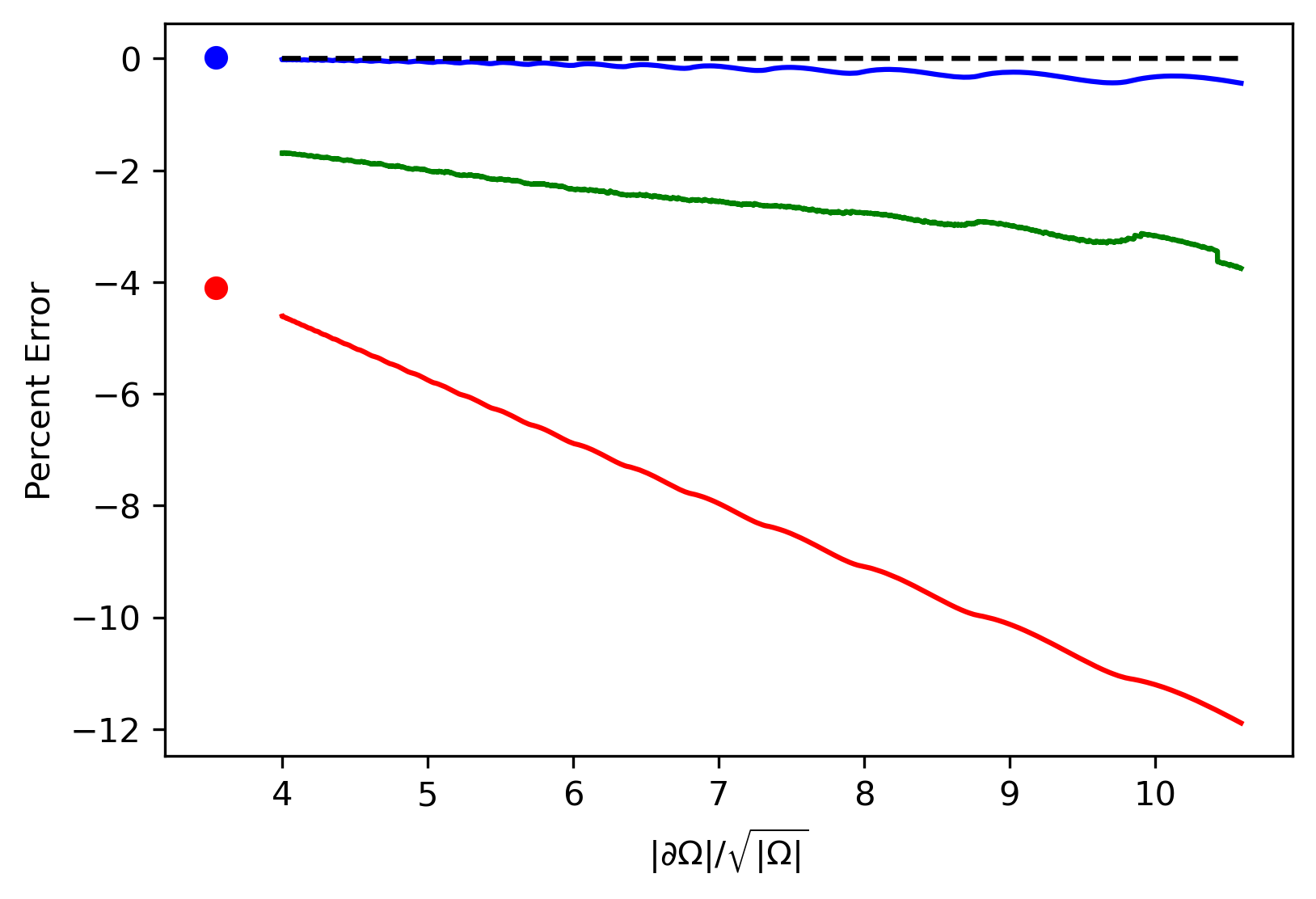}
\caption{Red is TF, green is TF on the exact density, and blue is ncTF, for a range of rectangular boxes (continuous) and for a circular cavity (discrete) at  $N=1000$. 
The dimensionless $|\partial \Omega|/\sqrt{|\Omega|}$ is proportionsl to $B$ (Tab. \ref{t:ab}) and equals $2\sqrt \pi$ for a circle and 4 for a square (see Secs~S6 and~S7 of \cite{SM}).
}\label{f:rectcomp}
\end{figure}

{\em Generality:
}
Table \ref{t:ab} presents further and richer examples of our method. In all cases, we find
\begin{equation}
  \ETF(N) =  A N^p,~~~~~~~  \Delta N = B N^q
  \label{Anp}
\end{equation}
A key feature is that the formula for the correction power $q$, unlike those for $p$, $A$, or $B$, is universal. Moreover, the sign of $B$ follows that of divergences in the potential, intuitively matching the error in the TF density in the interior.
Fig.~\ref{f:rectcomp} shows the significance of the normalization correction across cavities of different shapes. $B$ increases with aspect ratio, so TF correspondingly loses accuracy, while ncTF remains accurate even for very elongated boxes. The formula $\Delta N = (|\partial \Omega|/3 \sqrt{|\Omega|\pi})N^{1/2} $ works for 2D cavities of any shape,  with $|\Omega|$ being the area of the cavity and $|\partial\Omega|$ the perimeter. Table \ref{t:circular} lists results for a circular cavity, where $\Delta N = (2/3)N^{1/2}$.

\begin{table}[ht]
{\setlength{\extrarowheight}{3pt}
    \begin{tabular}{|c|c|c|c|}
    \hline
   $N$ &    $E(N)$ &    $\ETF(N)$  &  $\ENC(N) = \ETF(\NNC)$   \\ \hline
    19 & 487  & 361 (-26\%) & 480 (-2\%) \\
    30 & 1139 & 900 (-21.0\%) & 1132 (-0.6\%) \\
    100 & 11,408 & 10,000 (-12\%) & 11,378 (-0.3\%)\\
    1000 & 1,042,850 & 1,000,000 (-4\%) & 1,042,608 (-0.02\%)\\\hline
    \end{tabular}
}
\caption{Exact and approximate energies for a circular cavity of radius one. (Energy expressions in Sec.~S7 of \cite{SM})}\label{t:circular}
\end{table}

\begin{table*}[ht]

{\setlength{\extrarowheight}{10pt}
    \begin{tabular}{|c|c|c|c|c|c|c|}
    \hline
   \phantom{m}$d$\phantom{m}& Example
&Potential& \phantom{m}$p$\phantom{ml}  &$A$ &  $B$ & \phantom{mm} $q$ \phantom{mm}  \\[3pt]   \hline
    $1$ & interval of length $| \Omega|$&  $0$ on $(0,|\Omega|)$ and $+\infty$ elesewhere
  & $ 3$& $\displaystyle\frac{\pi^2}{6| \Omega|^2}$ & $\displaystyle\frac 12$ & $\displaystyle\frac {d-1} d$\\  [8pt] 
    \hline  
    $1$ & \parbox[t]{4cm}{\vspace{-4mm}  half harmonic oscillator \\ of frequency $\omega$} 
& $\frac 12 \omega x^2$ on $(0,\infty)$  and $+\infty$ elsewhere
& $2$& $  \omega$ & $\displaystyle \frac 14$ &$\displaystyle\frac {d-1} d$\\ [8pt] 
    \hline  
    $1$ & \parbox[t]{4cm}{\vspace{-4mm}  half linear well \\ of strength $F$} 
& $F x$ on $(0,\infty)$  and $+\infty$ elsewhere
& $\displaystyle \frac 53$& $ \displaystyle\frac {3(3 \pi F)^{\frac 23}}{10}$ & $\displaystyle \frac 14$ &$\displaystyle\frac {d-1} d$\\ [8pt] 
	\hline
    $2$ &    \parbox[t]{3cm}{\vspace{-4mm}  cavity of area $| \Omega|$\\ and perimeter $| \partial \Omega|$} & $0$ on $\Omega \subset \mathbb R^2$  and $+\infty$ elsewhere  & $ 2$  & $\displaystyle\frac \pi {| \Omega|}$ & $\displaystyle\frac {| \partial \Omega|}{3 \sqrt{| \Omega| \pi}}$ & $\displaystyle\frac {d-1} d$\\  [8pt] 
    \hline  
    $2$ &  \parbox[t]{4cm}{\vspace{-4mm}  quarter harmonic \\ oscillator of frequency $\omega$} 
  &\phantom{l} $ \frac 12 \omega r^2$ on $(0,\infty)\times(0,\infty)$ and $+\infty$ elsewhere \phantom{l}& $\displaystyle\frac 32$& $\displaystyle\frac {4 \sqrt 2}3 \omega$ & $\displaystyle \frac 1{2\sqrt 2}$ & $\displaystyle\frac {d-1} d$\\ [8pt] 
    \hline  
    $3$ & \parbox[t]{3cm}{\vspace{-4mm}  cavity of volume $| \Omega|$\\ and surface area $| \partial \Omega|$} &  $0$ on $\Omega \subset \mathbb R^3$ and $+\infty$ elsewhere & $\displaystyle\frac 53$ & $ \displaystyle\frac {3(6 \pi^2)^{\frac 23}}{10|\Omega|^{\frac 23}}$ & $ \displaystyle\frac{(36\pi)^{\frac 13}|\partial \Omega|}{32|\Omega|^{\frac 23}}$ & $\displaystyle\frac {d-1} d$\\ [8pt] 
\hline     $3$ &  Int. Coulomb, $Z=N$  & $ - \dfrac Z{r}$ on $\mathbb R^3 \setminus\{0\}$, $Z = N $
    & $\dfrac73$ & $-c_0$& $-\dfrac{3}{14 c_0}$ & $\displaystyle\frac {d-1} d$\\ [8pt]
    \hline     $3$ &  Non-int. Coulomb, $Z=N$  & $ - \dfrac Z{r}$ on $\mathbb R^3 \setminus\{0\}$, $Z = N $
    & $\dfrac73$ & $-\dfrac{3^{\frac13}}2$ & $-\dfrac{3^{\frac 23}}{14}$ & $\displaystyle\frac {d-1} d$\\ [8pt]
    \hline   
    $3$ &  Non-int. Coulomb, $Z$ fixed  & $ - \dfrac Z{r}$ on $\mathbb R^3 \setminus\{0\}$
    & $\dfrac13$ & $-\dfrac{3^{\frac13}}2$ & $-\dfrac{3^{\frac 23}}{2}$ & $\displaystyle\frac {d-1} d$\\ [8pt]
        \hline   
    \end{tabular}
}
\caption{ Constants in Eq. \ref{Anp}. The 1D results are from Eq. \eqref{e:1dderiv}, those for cavities from Weyl asymptotics, Eq.~\eqref{e:weyl}.  The quarter harmonic (Sections~S4 and~S5 of \cite{SM})  and Coulomb (Eq.\eqref{e:abpq}) results come from explicit formulas for eigenvalues.
}\label{t:ab}
\end{table*}

Generally, the more accurate an approximate energy functional is, the smaller $\Delta N$ is, but a good choice of $\Delta N$ improves energies ($\Delta N$ vanishes for the 1D harmonic oscillator because of perfect cancellation of errors in TF). Typically $\Delta N$ is small when the potential is smooth. For the two-dimensional isotropic harmonic oscillator $\Delta N = 1/24$ when all shells are filled, while the exact energy is
\begin{equation}\label{eq:harmos}
 E(N) = \frac {\omega}3 N\sqrt{8N+1}.   
\end{equation}
Neither of these examples fit in  Tab.~\ref{t:ab}, because in both cases the leading correction is zero.

We have given general formulas for $\Delta N$ in 1D, for any cavity in any dimension, and for specific cases.
A more general formula for  $\Delta N$ as a functional of the potential $v(\bf{r})$ would be very powerful and could come from a more general version of the Weyl asymptotics~\eqref{e:weyl}. Some relevant asymptotics have been computed for smooth potentials in \cite{HelRob-AA-90,GuiWan-JDG-12}, and the large body of work on the Scott correction, which corresponds to the Coulomb case, is discussed in \cite{FraMerSie-LMP-23}. It is natural to expect a formula involving a phase space integral, in the spirit of \cite{MarPla-MPS-56}.

{\em Interacting electrons:} In practical applications of DFT, electrons are subject to Coulomb repulsion.  In this many-body problem, there is a very specific semiclassical limit of all (non-relativistic) matter, in which the one-body potential is scaled along with $N$.  In the special case of neutral atoms, this corresponds
to simply keeping $Z=N$, where $Z$ is the number of protons in the nucleus.

Over many decades, the asymptotic expansion for neutral atoms was derived:
\[
 E(N) = - c_0 N^{7/3} + N^2/2  - c_2 N^{5/3}+ \cdots,
\]
where $c_0 = 0.769745\dots$ and $c_2 = 0.269900\dots$ \cite{Sco-PMJC-52, Sch-PRA-81, OkuBur-arxiv-21} and orbitals are doubly occupied.  Here, TF theory (including the Hartree approximation for the
electron-electron repulsion) yields precisely the first term alone, while 
the second is the Scott correction \cite{Sco-PMJC-52, Sch-PRA-80, OkuBur-arxiv-21}.   Setting $\Delta N= -3(14c_0)^{-1}N^{2/3}$ recovers the Scott correction, yielding the results in Table~\ref{t:noble}.
The asymptotic expansion of $\ENC$ yields a correction with $\tilde c_2 = 5/(28 c_0) = 0.232\dots$,  within $15\%$ of the correct value.

\begin{table}[ht]
{\setlength{\extrarowheight}{3pt}
    \begin{tabular}{|c|c|c|c|c|c|}
\hline
       \multicolumn{3}{|c}{} & \multicolumn{2}{|c|}{Percent errors}\\
\hline
   Atom & $N$ & $E(N)$ & $\ETF(N)$  
& $\ENC(N)$   \\
\hline
   He & $2$ & $-2.904$ & $-33$  
 & $26$   \\
   Ne & $10$ & $-128.9$ & $-28$  
 & $7$   \\
Ar & $18$ & $-527.6$ & $-24$ 
 & $5$ \\
Kr & $36$ & $-2754$ & $-19$ 
 & $2.8$ \\
Xe & $54$ & $-7235$ & $-17$ 
 & $2.1$ \\
Rn & $86$ & $-21870$ & $-15$ 
 & $1.5$ \\
\hline
    \end{tabular}
}
\caption{Energies and percent errors for TF 
and ncTF for nonrelativistic noble gases. 
For $N \ge 10$, $E(N)$ obtained by adding  exchange-only  energies from Table 4.6 of \cite{EngDre-BOOK-11} and acGGA+ correlation corrections from Table I of \cite{CanCheKruBur-JCP-18}; (correlation is unimportant, being less than  0.1 per electron).}\label{t:noble}
\end{table}

{\em The Bohr atom:} 
We can relate the interacting case above back to our non-interacting examples.
The Bohr atom~\cite{HeiLie-PRA-95,Eng-LNP-88, BurCanGouPit-JCP-16, OkuBur-arxiv-21} consists of non-interacting fermions (singly) occupying hydrogenic orbitals.  If the first $k$ shells are filled,   $N=1^2 + 2^2+ \cdots + k^2$, and $k(N)$ is the inverse function, then
\begin{equation}\label{e:abpq}
   E(N) = - \frac {Z^2} 2 k(N) =  - \frac {Z^2} 2\Big((3N)^{1/3} - \frac 12  + \cdots \Big)
\end{equation}

\begin{figure}[htb]
\includegraphics[width=6cm]{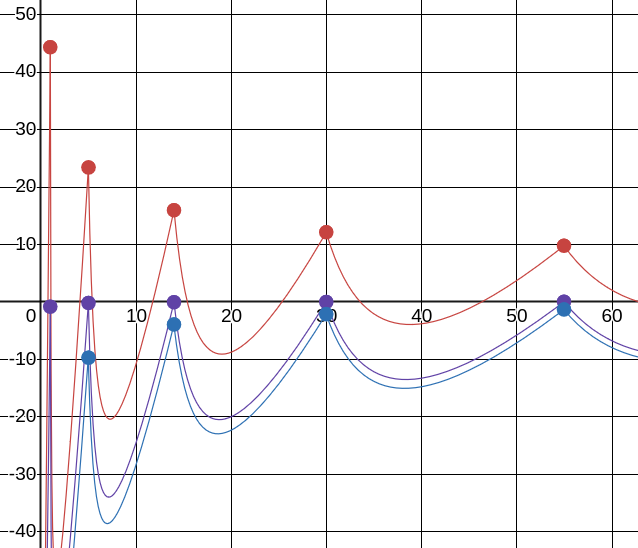}

\caption{
Percent errors in Bohr atom energies for TF (red), ncTF with $Z$ fixed (blue),
and ncTF with $Z=N$ (purple). See Section~S8 and Table~S2 of \cite{SM}.
}\label{f:coulomb}
\end{figure}

We consider two distinct expansions.   In the first,  $Z=1$ and  $N\gg1$, as in all our non-interacting examples.  In the second,  $Z=N\gg1$, as  for interacting problems.  
In Fig.~\ref{f:coulomb} (and Tab.~S2 of \cite{SM}), both significantly improve over simple TF, but the latter is more accurate, even more so than the Scott correction itself.

{\em Exchange-correlation(XC):}
The KS scheme approximates only a small portion of the energy, the XC energy.
The simplest approximation is LDAX, the local density approximation for exchange~\cite{Dir-MPCPS-30,Blo-ZfP-29,Sla-PR-51},
which underestimates its magnitude~\cite{Esc-BOOK-96}. It uses the exchange energy density of electrons in an infinite box, analogously to the TF calculation for Fig.~\ref{f:pboxintro}.  We take an optimistic leap and imagine 
that the same forms of Table III apply to LDAX, and simply multiply the density by a factor $1+\Delta N/N$, with $\Delta N = B N^{2/3}$. 
This yields the normalization-corrected value
\begin{equation}
    E\x\LDA({\rm NC}) = (1+B/N^{1/3})^{4/3} E\x\LDA.
\end{equation}
Choosing $B=0.125$ by eye,
we find the remarkable improvements shown in Table~\ref{t:Ex} for the noble gas atoms, mirroring those for the total energy. The work of 
\cite{CorsoFriesecke} {\em derives} the exact same form for the exchange asymptotics in a 3D box, with $B \approx 0.47$ (see Section S9 of \cite{SM}), consistent with our guess. Their kinetic energy
calculation agrees with our formulas, and also yields a bulk density
of $N+\Delta N$ per unit volume, just as in our examples.

\begin{table}[t!]
{\setlength{\extrarowheight}{3pt}
    \begin{tabular}{c|c|c|c|c|c}
   Atom & $N$ &$\left|{\rm{EXX}(N)}\right|$& $\left|E^{\rm LDA}_{\rm x}(N)\right|$ & $\left|E^{\rm PBE}_{\rm x}(N)\right|$  & $\left|E^{\rm LDA}_{\rm nc}(N)\right|$\\
    \hline
He &2 & 1.026 & $0.862 \  (16\%)$ & $1.005 \  (2\%)$ & $0.978 \ (5\%)$\\
Ne & 10 & 12.10 & $10.97 \  (9\%)$ & $12.03 \  (0.6\%)$ &  $11.82 \ (2\%)$\\
Ar & 18 & 30.18 & $27.81 \  (8\%)$ & $29.98 \  (0.64\%)$&  $29.59 \ (2\%)$\\
Kr & 36 & 93.83 & $88.53 \  (6\%)$ & $93.38 \  (0.5\%)$  & $93.03 \ (0.9\%)$ \\
Xe & 54 & 179.1 & $170.5 \  (5\%)$ & $178.2 \  (0.5\%)$ &  $178.1 \ (0.5\%)$\\
Rn & 86 & 387.5 & $372.9 \  (4\%)$ & $385.9 \  (0.4\%)$ &  $387.0 \ (0.1\%)$ \\
    \end{tabular}
}
\caption{  Energies (percentage errors) for the noble gases with EXX$(N)$ being the exact exchange energy, $E^{\rm LDA}_{\rm x}(N)$  the LDA exchange energy, $E^{\rm PBE}_{\rm x}(N)$ the PBE exchange energy, and $E^{\rm LDA}_{\rm nc}(N)$ the normalization-corrected LDA. The first three were obtained from \cite{CanCheKruBur-JCP-18}.}\label{t:Ex}
\end{table}

{\em Outlook:}
For a one-dimensional box, one can compute $\Delta N$ by eye from  Fig.~\ref{f:pboxintro} by making the straight line of the approximate density go through the middle of the oscillations of the exact density. In less obvious cases, the practicality of normalization corrections will ultimately rely on more robust formulas for $\Delta N$  as a functional of the potential.
The above derivations based on WKB and Weyl asymptotics are special cases of semiclassical asymptotics, which admit many general formulations \cite{dimassisjostrand,zworski,guilleminsternberg, ivrii100}. Work deriving $\Delta N$ in a  universal way from semiclassical trace formulas \cite{HelRob-AA-90,GuiWan-JDG-12}, building on the phase space point of view of \cite{MarPla-MPS-56, CanLeeEllBur-PRB-10}, is ongoing.   Our work emphasizes the need for results with open boundaries rather than cavities.  Such results could lead to improvements in accuracy for orbital-free DFT calculations, which scale linearly in $N$, avoiding the cubic-in-$N$ calculation of Kohn--Sham orbitals~\cite{MiLuoTriPav-CR-23}. For applications to solids, these results should generalize to periodic boundary conditions.  
 Thus, we make DFT accurate up to two orders instead of just the leading one in a way analogous to -- but different from -- corrections that were used to construct PBEsol~\cite{PerRuzCsoVydScuConZhoBur-PRL-08} and SCAN~\cite{SunRuzPer-PRL-15}.

The density with $\Delta N$ extra electrons is not the true density of the system; it is a device guaranteeing more accurate energetics by improving asymptotics.  In principle, one could  derive a correction to the TF density functional whose optimal density reproduces these energetics, by taking the functional derivative of the corrected energy with respect to the potential. 
In practice, this is a non-trivial   derivation which likely requires some form of regularization in the vicinity of the boundaries. 

In future, there are many variations of our tricks that could be applied to the XC functional in a KS calculation, not just the one tried here. 
 Important open problems include proving that our method is  applicable to the XC energy, making it size-extensive, and using it to guide the development of better functionals. 

Our analysis is also relevant to 
improving {\em density-corrected DFT} (DC-DFT), a simple method that has become popular
 \cite{SimSonVucBur-JACS-22,VucSonKozSimBur-JCTC-19,SonVucSimBur-JCTC-22}, where the self-consistent density is replaced by (a proxy for) the exact density. 
DC-DFT has proven successful in many applications, including  ions in solution~\cite{KimSimBur-JCP-14} and the phase diagram of water~\cite{SonVucKimYuSimBur-Nat-23}. 
But we have shown that a normalization correction of an approximate density can be more accurate than evaluation on the exact density.  Our correction is also much easier to implement, not just in this example but in every case we have considered.

{\em Acknowledgments:} K.D. was supported by NSF Award No. DMS-1708511  and by a  Simons Foundation Collaboration Grant for Mathematicians.  W.M. and A.W. were supported by NSF Award No. CHE-2306011.  K.J.D. would like to thank UCI's Chancellors Postdoctoral Fellowship Program and in particular Prof. Dr. Feizal Waffarn for his support as a sponsor. 
 K.B. was supported by the NSF Award No. CHE-2154371. Thanks also to Hamid Hezari and Antoine Prouff for helpful discussions, and to Vienna Cafe for warm hospitality. K.B. also thanks Gero Friesecke, and Thiago  Carvalho Corso, and the Mathematisches Forschungsinstitut Oberwolfach for hospitality.

\bibliography{Bibliography}

\end{document}


\title{Supplemental Material: Approximate normalizations for approximate density functionals}

\author{Adam Clay$^1$}
\author{Kiril Datchev$^1$}
\author{Wenlan Miao$^2$}
\author{Adam Wasserman$^{2,3}$}
\author{Kimberly J. Daas$^4$}
\author{Kieron Burke$^{4,5}$}
\affiliation{$^1$Department of Mathematics, Purdue University, West Lafayette, IN 47907, USA}

\affiliation{$^2$Department of Physics and Astronomy, Purdue University, West Lafayette, IN 47907, USA}

\affiliation{$^3$Department of Chemistry, Purdue University, West Lafayette, IN 47907, USA} 

\affiliation{$^4$Department of Chemistry, University of California, Irvine, CA 92697, USA}

\affiliation{$^5$Department of Physics and Astronomy, University of California, Irvine, CA 92697, USA}

\maketitle
\tableofcontents
\section{Energy expressions of the one-dimensional box}\label{SI:1DBox}
The exact energy of $N$ particles in a box of length $L$ is
\begin{equation*}
    E\left(N\right)=\frac{\pi^{2}}{6L^{2}}\left(N^3+\frac{3}{2}N^2+\frac{1}{2}N\right).
\end{equation*}
The TF energy in this case is
\begin{equation*}
    \ETF\left(N\right)=\frac{\pi^{2}}{6L^{2}}N^{3}
\end{equation*}
and the normalization corrected energy is
\begin{equation*}
    \ENC\left(N\right)=\frac{\pi^{2}}{6L^{2}}\left(N+\frac{1}{2}\right)^{3}=\frac{\pi^{2}}{6L^{2}}\left(N^3+\frac{3}{2}N^2+\frac{3}{4}N+\frac{1}{8}\right).
\end{equation*}
The exact density in the TF energy functional gives,
\begin{equation*}
\widetilde{E}_{\rm d}\left(N\right)=\frac{4\pi^{2}}{3L^3}\int_{0}^{L}\left(\sum_{n=1}^{N}\sin^2\left(\frac{n\pi x}{L}\right)\right)^{3}dx = \frac{\pi^{2}}{3L^3}\int_{0}^{L}\left(2N+1- \frac{\sin((2N+1) \pi x/L)}{\sin(\pi x/L)}\right)^{3}dx,
\end{equation*}
which simplifies to
\begin{equation*}
\widetilde{E}_{\rm d}\left(N\right)=\frac{\pi^2}{6L^2}\left(N^3+\frac{9}{8}N^2+\frac{3}{8}N\right),
\end{equation*}
as shown in  Ref.~\cite{EllLeeCanBur-PRL-08} with $\zeta=1$.

\noindent

 In this example it is easy to compute the value of $\Delta N$ which gives the exact result $\ETF(N + \Delta N) = E(N)$. A simple algebraic calculation yields
\[
\Delta N = \sqrt[3]{N^3+\frac{3}{2}N^2+\frac{1}{2}N} - N = \frac 12 - \frac 1 {12 N} + \cdots,
\]
confirming that $1/2$ is the leading correction. A simpler formula which captures the next three terms is
\begin{equation} \label{e:super}
\Delta N = \frac 12 - \frac 1{12} \ln\Big(1 + \frac 1 {N}\Big).
\end{equation}

\begin{figure}[h]
\includegraphics[width=9cm]{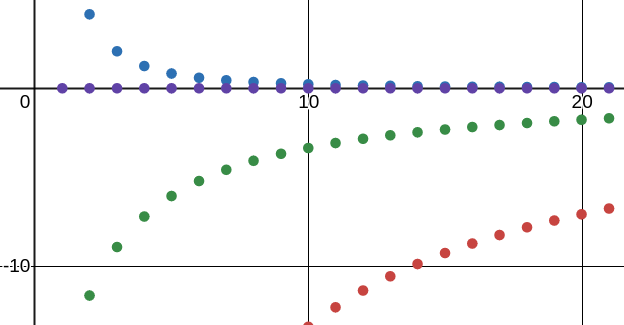} \hspace{.5cm} \includegraphics[width=7cm]{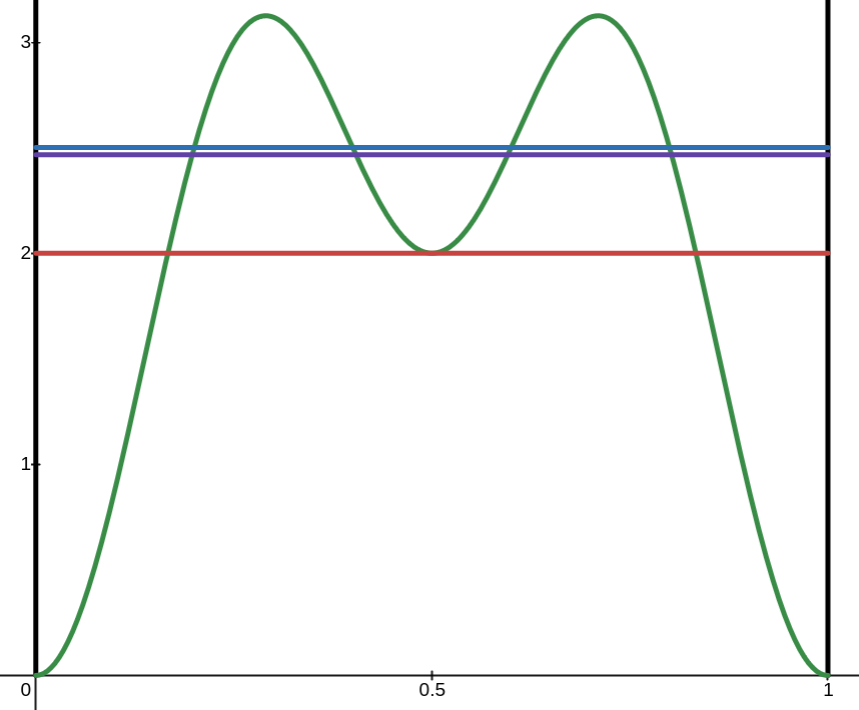}
\caption{The same plots as Figure \ref{f:pboxintro}, with the improved correction from \eqref{e:super} in purple. The percent error for purple is  $-10^{-4}$ when $N=2$, $-10^{-8}$ when $N=11$, and continues to improve from there.}
\end{figure}
{\em Ionization energies:} 
The normalization-correction approach also leads to improved accuracy for ionization energies. For example, for $N$ non-interacting electrons in a one-dimensional box of length $L$, the exact ionization energy is 
\begin{equation}
\nonumber
I=E(N)-E(N-1)=\frac{\pi^2}{2L^2}N^2
\end{equation}
The Thomas-Fermi approximation is
\begin{eqnarray}
\nonumber
\widetilde{I}&=&\widetilde{E}(N)-\widetilde{E}(N-1)\\
&=&\frac{\pi^2}{2L^2}\left(N^2-N+\frac{1}{3}\right)
\end{eqnarray}
The linear-in-$N$ error is eliminated by our normalization-correction approach, since:
\begin{eqnarray}
\nonumber
\widetilde{I}_{\textrm{nc}}&=&\widetilde{E}(N+\frac{1}{2})-\widetilde{E}(N-\frac{1}{2})\\
&=&\frac{\pi^2}{2L^2}\left(N^2+\frac{1}{12}\right)
\end{eqnarray}

\section{Derivation of general energy expressions in one dimension}\label{SI:1Dgen}  To derive Eq.~\eqref{e:1dexact}, observe that 
\[
1-s'(\alpha) = \sum_{j=-\infty}^\infty \delta(\alpha-j).
\]
Hence
\[
 E(N) = \sum_{j=1}^N \mathcal{E}(\lambda(j))= \int_{1/2}^{N+1/2} d\alpha \, \Big(\mathcal{E}(\lambda(\alpha))(1-s'(\alpha))\Big).
\]
Now integrate by parts in the second term to obtain
\[
 E(N) = \int_{1/2}^{N+1/2} d\alpha \, \Big(\mathcal{E}(\lambda(\alpha))+ \mathcal {E}'(\lambda(\alpha))\lambda'(\alpha)s(\alpha)\Big),
\]
where we  used $s(1/2) = s(N+1/2) = 0$. This is Eq.~\eqref{e:1dexact}.

\begin{figure}[h]
\includegraphics[width=10cm]{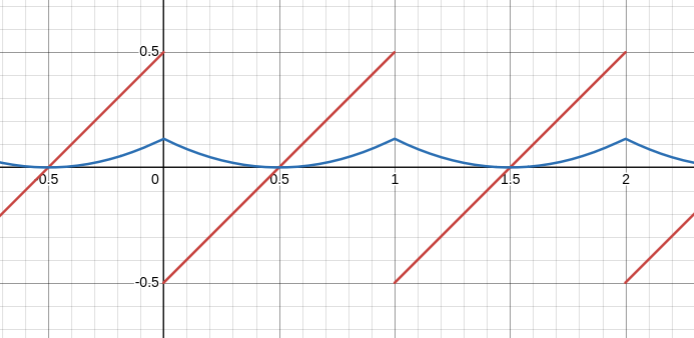}
\caption{The auxiliary functions $s$ is in red and $S$ in blue, where $S$ is an antiderivative of $s$.}
\end{figure}

Getting from Eq.~\eqref{e:1dexact} to Eq.~\eqref{e:1dderiv} is much more technical. We begin with an outline of the argument.

First, we will show that writing
\[
\int_{1/2}^{N+1/2} d\alpha \, \Big(\mathcal{E}(\lambda(\alpha))+ \mathcal {E}'(\lambda(\alpha))\lambda'(\alpha)s(\alpha)\Big) \approx \int_{1/2}^{N+1/2} d\alpha  \, \mathcal{E}(\lambda(\alpha))
\]
is correct because the integral of $\mathcal{E}'(\lambda(\alpha)) \lambda'(\alpha) s(\alpha)$ is two orders smaller than the integral of $\mathcal{E}(\lambda(\alpha))$, one order being gained by going from $\mathcal E$ to $\mathcal E'$, and another order being gained by the oscillations of $s$, which are exploited using integration by parts. 

Next, we will show that writing
\[
\int_{1/2}^{N+1/2} d\alpha  \, \mathcal{E}(\lambda(\alpha)) \approx \int_{1/2}^{N+1/2} d\alpha  \, \mathcal{E}(\alpha - \nu) = \int_{-\nu+1/2}^{N-\nu+1/2} d\lambda  \, \mathcal{E}(\lambda)
\]
is correct because $\lambda(\alpha) - (\alpha - \nu)$ is two orders smaller than $\lambda(\alpha)$. 
Finally, we will deduce \eqref{e:1dderiv} by showing that writing 
\[
 \int_{-\nu+1/2}^{N-\nu+1/2} d\lambda  \, \mathcal{E}(\lambda) \approx \int_{0}^{N-\nu+1/2} d\lambda  \, \mathcal{E}(\lambda) 
\]
is correct because the integral from $0$ to $-\nu+1/2$ is bounded in $N$.

To justify these steps we make the following assumptions. Assume:
\[
\mathcal E(\lambda) = O(\lambda^{p-1}), \ \mathcal E'(\lambda) = O(\lambda^{p-2}), \ \mathcal E''(\lambda) = O(\lambda^{p-3}),
\]
\[
 \lambda(\alpha) = O(\alpha), \ \lambda'(\alpha) = O(1), \ \lambda''(\alpha) = O(\alpha^{-1}),
\]
\[
 r(\alpha) = O(\alpha^{-1}), \  r'(\alpha) = O(\alpha^{-2})
\]
\[
 \mathcal E(\lambda)^{-1} = O(\lambda^{1-p}).
\]

To bound the second term of Eq.~\eqref{e:1dexact}, integrate by parts again to get
\[
  \int_{1/2}^{N+1/2} d\alpha\,  \mathcal {E}'(\lambda(\alpha))\lambda'(\alpha)s(\alpha) =
 -  \int_{1/2}^{N+1/2} d\alpha \, \Big(\mathcal {E}'(\lambda(\alpha))\lambda'(\alpha)\Big)'S(\alpha),
\]
where $S(\lambda) = \int_{1/2}^\lambda s$; note that $S$ vanishes at all the half-integers.  

Then use
\[
 |(\mathcal {E}''(\lambda(\alpha))\lambda'(\alpha)^2 + \mathcal {E}'(\lambda(\alpha))\lambda''(\alpha))S(\alpha)| = O(\alpha^{p-3}),
\]
to get 
\[
\Big|  E(N) - \int_{1/2}^{N+1/2} d\alpha \, \mathcal{E}(\lambda(\alpha)) \Big| = O(N^{p-2}).
\]
Next, use $\lambda'(\alpha) = 1 + r'(\alpha)$ to write, assuming that $\lambda$ is extended from integer to noninteger values so as to be increasing,
\[
 \int_{1/2}^{N+1/2} d\alpha \, \mathcal{E}(\lambda(\alpha))= \int_{1/2-\nu+r(1/2)}^{N+1/2-\nu+r(N+1/2)} \frac{d\lambda\, \mathcal E(\lambda)}{1+r'(\alpha(\lambda))}.
\]
Since
\[
 \frac 1 {1+r'(\alpha(\lambda))} = 1 -  \frac {r'(\alpha(\lambda))} {1+r'(\alpha(\lambda))},
\]
and $r'(\alpha(\lambda)) = O(\lambda^{-2})$, we see that
\[
 \frac{\mathcal E(\lambda) r'(\alpha(\lambda))}{1 +  r'(\alpha(\lambda))} = \frac{O(\lambda^{p-1}) O(\lambda^{-2})}{1 + O(\lambda^{-2})} = O(\lambda^{p-3}),
\]
so that
\[
 \Big|  \int_{1/2-\nu+r(1/2)}^{N+1/2-\nu+r(N+1/2)} d\lambda \, \Big(\frac{ \mathcal E(\lambda)}{1+r'(\alpha(\lambda))} -  \mathcal E(\lambda) \Big)\Big| = O(N^{p-2}).
\]
Meanwhile
\[
  \Big| \int_{1/2-\nu+r(1/2)}^{N+1/2-\nu+r(N+1/2)}d\lambda\, \mathcal E(\lambda)  - \int_{1/2-\nu+r(1/2)}^{N+1/2-\nu}d\lambda\, \mathcal E(\lambda)  \Big| = O(N^{p-2}),
\]
because the difference is integrating a function of size $O(N^{p-1})$ over a region of size $O(N^{-1})$.
Similarly, we can adjust the lower limit:
\[
\Big| \int_{1/2-\nu+r(1/2)}^{N+1/2-\nu}d\lambda\, \mathcal E(\lambda)  - \int_{0}^{N+1/2-\nu}d\lambda\, \mathcal E(\lambda)  \Big| = O(1).
\]
Putting together these bounds yields
\[
 \Big |E(N) - \int_{0}^{N+1/2-\nu}d\lambda\, \mathcal E(\lambda)  \Big| = O(N^{p-2}) + O(1).
\]
If $p > 2$ then  the first remainder dominates, and if $p<2$ then the second one does.

  We remark that there always exists  $\Delta N$ such that $\ETF(N + \Delta N) = E(N)$ exactly. This is because  
\[
 \int_0^{N + \Delta N} d \lambda \,\mathcal E(\lambda) = \ETF(N + \Delta N)
 \]
 for any $\Delta N$, since $p(\mathcal E,x)/\pi$ is always the Thomas--Fermi density. Next, adding a constant to $v(x)$ we may arrange that $\mathcal E(\lambda) \ge 0$ for all $\lambda$, and then  $\ETF(N + \Delta N) \to \infty$ as $\Delta N \to \infty,$ while $\ETF(0) = 0$, so there must exist $\Delta N$ for which $\ETF(N + \Delta N) $ passes through the value $E(N)$. An example is given in Section \ref{SI:1DBox}.

\section{Energy expressions of the three-dimensional rectangular box} \label{SI:3Dbox}
A 3d rectangular box with sides lengths $L_x$, $L_y$ and $L_z$ has energy levels
\begin{equation*}
\begin{aligned}
&E(N)=\sum_{i,j,k}^{E_{ijk}<R(N)^2}\mathcal E_{ijk}\\
&\mathcal E_{ijk}=\frac{1}{2}\frac{i^2\pi^2}{L_x^2}+\frac{1}{2}\frac{j^2\pi^2}{L_y^2}+\frac{1}{2}\frac{k^2\pi^2}{L_z^2},
\end{aligned}
\end{equation*}
where $R(N)$ is the radius of the quantum number space:
$R(N)=(2^3N/(L_xL_yL_z))^{1/3}$. The TF energy is
\begin{equation*}
    \ETF(N)=\frac{3(6\pi^2)^{\frac{2}{3}}}{10|\Omega|^{\frac{2}{3}}}N^\frac{5}{3}
\end{equation*}
while the normalization corrected energy is  
\begin{equation*}
    \ENC(N)=\frac{3(6\pi^2)^{\frac{2}{3}}}{10|\Omega|^{\frac{2}{3}}}\left(N+\frac{(36\pi^\frac{1}{3})|\partial \Omega|}{32|\Omega|^{\frac{2}{3}}}N^\frac{2}{3}\right)^{\frac{5}{3}}.
\end{equation*}
The TF energy on the exact density is
\begin{widetext}
\begin{equation*}
\widetilde{E}_{\rm d}(N)=\frac{24(6\pi^2)^{\frac{2}{3}}}{10|{\Omega}|}\iint \sum_{i,j,k}^{E_{ijk}<R(N)^2} \left (\sin^2\left(\frac{i\pi x}{L_x}\right)\sin^2\left(\frac{j\pi y}{L_y}\right)\sin^2\left(\frac{k\pi z}{L_z^2}\right)\right )^{\frac{5}{3}}dxdydz.
\end{equation*}
\end{widetext}
The energies and percentage errors from $N=1$ to $N=1000$ can be found in 3d\_box.txt.

\section{Energy expressions of the two-dimensional Harmonic Oscillator}\label{SI:2DHO}  
For the two-dimensional harmonic oscillator with frequency $\omega$ the exact energy expression is given in Eq.~\ref{eq:harmos} of the main text. To derive it, let $H = - (1/2) \nabla^2 + \frac 12 \omega^2 r^2$. The energy levels are
\[
 \mathcal E_{m,n} = (m+n+1)\omega,
\]
where $m$ and $n$ are the quantum numbers corresponding to the $x$ and $y$ space coordinates.
Hence the energy level $i \omega$ has degeneracy $i$. If we have $N$ particles filling up energy levels up to the $N_\ell$th level, then 
\[
 N =  \sum_{i=1}^{N_\ell} i = \frac 1 2(N_\ell^2 + N_\ell), \qquad \text{or} \qquad N_\ell = \frac 12 \sqrt{8N+1} - \frac 12.
\]
The total energy is
\[
 E(N) = \sum_{i=1}^{N_\ell}  i \cdot i \omega = N_\ell(N_\ell+1)(2 N_\ell+1) \frac \omega 6= 
 \frac {\omega}3 N\sqrt{8N+1} =   \frac{2\sqrt 2}{3} \omega N^{3/2} + \frac 1 {12\sqrt2} \omega N^{1/2} + \cdots,
\] 
the TF energy  is
\begin{equation*}
\ETF\left(N\right)= \frac{2\sqrt 2}{3} \omega N^{3/2},
\end{equation*}
and the normalization corrected energy is
\begin{equation*}
  \ENC\left(N\right)= \frac{2\sqrt 2}{3} \omega \Big(N + \frac 1 {24} \Big)^{3/2}.
\end{equation*}

\section{Energy expressions of the two-dimensional quarter Harmonic Oscillator}\label{SI:2DqHO}
The quarter harmonic oscillator is similar. In that case, the energy levels are
\[
 \mathcal  E_{m,n} = (2m+2n+3)\omega.
\]
which means the energy level $(2i+1) \omega$ has degeneracy $i$. 
 If we have $N$ particles filling up energy levels up to the $N_\ell$th level, then 
\[
 N =  \sum_{i=1}^{N_\ell} i = \frac 12 (N_\ell^2 + N_\ell), \qquad \text{or} \qquad N_\ell = \tfrac 12 \sqrt{8N+1} - \tfrac 12.
\]
The total energy is
\[
 E(N) = \sum_{i=1}^{N_\ell}  i \cdot (2i+1) \omega = N_\ell(N_\ell+1)(4 N_\ell+5) \frac \omega 6 =  N \Big(2 \sqrt{8N+1} + 3\Big) \frac \omega 3 = \frac {4\sqrt 2}3 \omega N^{3/2}+ \omega N + \cdots,
\]
the TF energy is
\begin{equation*}
\ETF\left(N\right)= \frac{4\sqrt 2}{3} \omega N^{3/2},
\end{equation*}
and the normalization corrected energy is
\begin{equation*}
  \ENC\left(N\right)= \frac{4\sqrt 2}{3} \omega \Big(N + \frac 1 {2\sqrt2} N^{1/2} \Big)^{3/2}.
\end{equation*}

\section{Energy expressions of the two-dimensional rectangular box}\label{SI:2DBox} 
A rectangular box with side lengths $L_x$ and $L_y$ has energy levels
\[
\mathcal E_{i,j} = \frac{1}{2}\frac{i^2\pi^2}{L_x^2}+\frac{1}{2}\frac{j^2\pi^2}{L_y^2},
\]
where  $i$ and $j$ are the quantum numbers corresponding to the sides of length $L_x$ and $L_y$ respectively.
The exact energy is given by,
\begin{equation*}
E(N)=\sum_{i,j}^{\mathcal E_{ij}<R(N)^2}\mathcal E_{ij},
\end{equation*}
where $R(N)$ is the radius of the quantum number space: $R(N)=(2^2N/(L_xL_y))^{\frac{1}{2}}$. The TF energy is
\begin{equation*}
    \ETF(N)=AN^p=\frac{\pi}{|\Omega|}N^2,
\end{equation*}
and the normalization corrected energy is
\begin{equation*}
 \ENC(N)=\frac{\pi}{|\Omega|}\left(N+\frac{|\partial \Omega|}{3\sqrt{|\Omega| \pi}} N^{1/2}\right)^2,
\end{equation*}
because the Weyl asymptotic formula is
\[
E(N) = \frac \pi A N^2 + \frac{2\sqrt \pi |\partial \Omega|}{3  |\Omega|^{3/2}} N^{3/2} + \cdots 
\]
The TF energy with the exact density can be written as
\begin{equation*}
\tilde{E}_d(N)=\frac{4\pi}{|\Omega|}\iint \sum_{i,j}^{E_{i,j}<R(N)^2}  \left(\sin^2\left(\frac{i\pi x}{L_x}\right)\sin^2\left(\frac{j\pi y}{L_y}\right)\right)^2\ dx dy.
\end{equation*}.
The energies and percentage errors for $N=1000$ can be found in 2d\_box.txt.

\begin{figure}
\includegraphics[width=14cm]{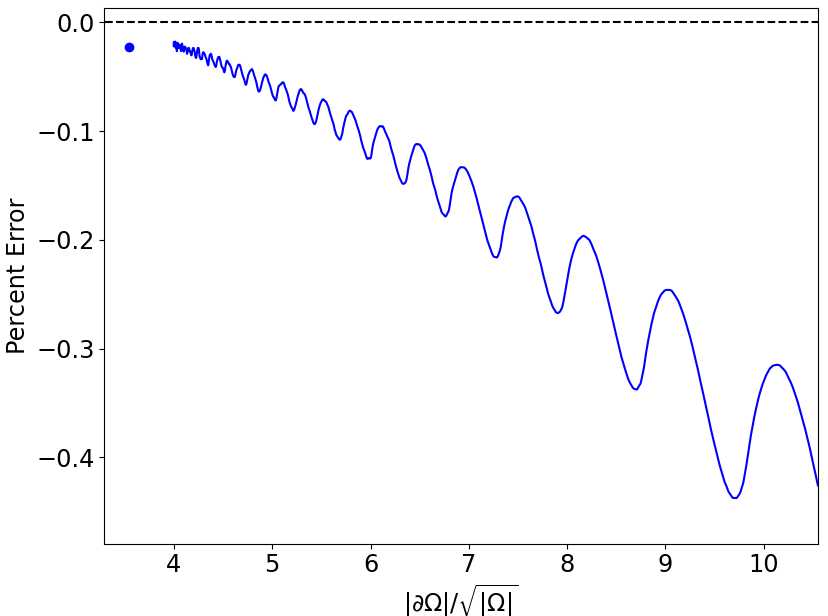}
\caption{A zoom of  Figure \ref{f:rectcomp}.}
\end{figure}

\section{Energy expressions of the two-dimensional circular cavity}\label{SI:2Dcav}

Separation of variables yields the radial equation
\[
 - \frac 12 u''(r) - \frac 1{2r} u'(r) + \frac {\ell^2}{2r^2} u(r) = \mathcal E  u(r),
\]
 where $\ell = 0,1,2, \dots$ is the angular momentum. Solutions which are regular at $r=0$ are given in terms of Bessel functions by $u(r) = J_\ell(\sqrt{2\mathcal E}r)$. Imposing the boundary condition $u(R) = 0$ tells us that, if the cavity has radius $R$, then $\sqrt{2\mathcal E}R$ must be a zero of the Bessel function $J_\ell$. In other words, the energy levels are
\[
\mathcal E_{\ell,n} =  \frac{j_{\ell,n}^2}{2R^2},
\]
where $j_{\ell,n}$ is the $n$th zero of $J_\ell$. Thus, if $R=1$, then the first twelve energy levels in order are
\[
 \mathcal E_{0,1} \approx 2.89, \quad  \mathcal E_{1,1} \approx 7.34, \quad  \mathcal E_{2,1} \approx 13.19, \quad  \mathcal E_{0,2} \approx 15.24, \quad \mathcal E_{3,1} \approx 20.35, \quad \mathcal E_{1,2} \approx 24.61, 
\]
\[
 \mathcal E_{4,1} \approx 28.79, \quad  \mathcal E_{2,2} \approx 35.42, \quad  \mathcal E_{0,3} \approx 37.43, \quad \mathcal E_{5,1} \approx 38.47, \quad \mathcal E_{3,2} \approx 47.64, \quad \mathcal E_{6,1} \approx 49.36, 
\]
Energy levels with $\ell = 0$ are nondegenerate, because the angular wave function is constant, while those with $\ell \ge 1$ are doubly degenerate, because the angular wave function is a combination of $\cos(\ell \theta)$ and $\sin (\ell \theta)$; there are no further degeneracies because $j_{\ell,n}\ne j_{\ell'n'}$ when $(\ell,n) \ne (\ell',n')$ \cite[\S 15.28]{watson}. The energies and percentage errors for $N=1000$ can be found in 2d\_box.txt. The exact density is given in terms of Bessel functions by
\begin{equation*}
n(\textbf{r})
\;=\;
\frac{1}{\pi}
\sum_{\{m,k\}\in \text{lowest }N}
c_{mk} 
\,
\bigl[
J_{m}\bigl(j_{mk}r\bigr)
\bigr]^{2}.
\end{equation*}
 with
\begin{equation*}
c_{mk}  \;=\; \frac{g_{mk}}{\bigl[J_{m-1}(j_{mk})\bigr]^{2}},
\end{equation*}
where $g_{mk}$ are the occupation numbers for each state:
\begin{equation*} g_{mk}=
\begin{cases} 1,\quad m=0,\\2,\quad m\ge1.
\end{cases}
\end{equation*}
The TF density is
\begin{equation*}
n^{\rm TF}(\textbf{r})=\frac{N}{\pi  },
\end{equation*}
and the normalization corrected density is
\begin{equation*}
n^{\rm TF}_{\rm nc}(\textbf{r})=\frac{N+\sqrt{N}}{\pi  }.
\end{equation*}
The densities on a grid from $r=0$ to $r=1$ can be found in 2d\_cavity.txt.

\section{Energy expressions for the Bohr atoms}\label{SI:Bohr}
The exact energies for the closed shell atoms are
\begin{equation*}
    E(N)=-\frac{N^2}{4}\left(A(N)-1+\frac{1}{3A(N)}\right)
\end{equation*}
with $A(N)^3=12 N \left(1-\sqrt{1-(3888 N^2)^{-1}}\right)$, which is a version of the formula from \cite{OkuBur-arxiv-21}, modified for singly occupied orbitals. For open shell atoms the formula is,
\begin{equation*}
    E(N)=-\frac{N^2}{2}\sum^{N}_{n=1} f_n
\end{equation*}
where $f_n$ is the $n$th element of the series of $\frac{1}{n^2}$ where each member is repeated $n^2$ times, i.e.
\begin{equation*}
    1,\frac{1}{4},\frac{1}{4},\frac{1}{4},\frac{1}{4},\frac{1}{9},\ldots,\frac{1}{9},\frac{1}{16},\ldots,\frac{1}{16},\frac{1}{25},\ldots,\frac{1}{25},\ldots;
\end{equation*}
see \cite{BurCanGouPit-JCP-16} for more details. The TF energy is \cite{HeiLie-PRA-95,OkuBur-arxiv-21},
\begin{equation*}
    \ETF(N)=-\dfrac{3^{\frac13}}2N^{\frac{7}{3}}.
\end{equation*}
If the Scott correction is added, one obtains
\begin{equation*}
    \widetilde{E}_{\rm sc}(N)=-\dfrac{3^{\frac13}}2N^{\frac{7}{3}}+\frac{1}{4}N^2,
\end{equation*}
where again the extra factor $\frac{1}{2}$ in front of the Scott correction comes from singly occupying the orbitals.
The resulting fixed $Z$ ($Z=1$ and $N\gg1$) normalization corrections are given by
\begin{equation*}
    \ETF(N)=-\dfrac{3^{\frac13}}2\left(N-\dfrac{3^{\frac23}}2N^{\frac{2}{3}}\right)^{\frac{7}{3}},
\end{equation*}
whereas for $N=Z\gg1$ they can be found via
\begin{equation*}
        \ENC(N)=-\dfrac{3^{\frac13}}2\left(N-\dfrac{3^{\frac23}}{14}N^{\frac{2}{3}}\right)^{\frac{7}{3}}.
\end{equation*}

\section{Normalization correction for LDAX for a 3D Box}\label{s:cf}

By Theorem 1.1 of \cite{CorsoFriesecke},
\[
 E^{LDA}_x(N) = - c_x \tilde \rho^{4/3} |\Omega| L^3 - c^{Dir}_{x,2} \tilde \rho |\partial \Omega| L^2 + \cdots,
\]
where $c_x = (3/4)(3/\pi)^{1/3}$, $c^{Dir}_{x,2} \approx 0.0767$, $\tilde \rho = N/(|\Omega|L^3)$, and $L$ is a scaling parameter for the size of the box. Absorbing $L^3$ into $|\Omega|$ and $L^2$ into $|\partial \Omega|$ yields
\[
  E^{LDA}_x(N) = - \frac{c_x}{|\Omega|^{1/3}} N^{4/3} - \frac{c^{Dir}_{x,2}|\partial \Omega|}{|\Omega|} N + \cdots.
\]
Next,
\[
 - \frac{c_x}{|\Omega|^{1/3}} (N + \Delta N)^{4/3} =  - \frac{c_x}{|\Omega|^{1/3}} N^{4/3} -  \frac{4c_x}{3|\Omega|^{1/3}} N^{1/3} \Delta N + \cdots,
\]
Matching the second term on the right side of each equation yields
\[
 \frac{c^{Dir}_{x,2}|\partial \Omega|}{|\Omega|} N =  \frac{4c_x}{3|\Omega|^{1/3}} N^{1/3} \Delta N,
\]
or
\[
 \Delta N =  B N^q, \qquad  B = \frac{3c^{Dir}_{x,2}|\partial \Omega|}{4c_x|\Omega|^{2/3}}, \qquad q =\frac 23.
\]
Comparing with Table \ref{t:ab}, we see that $q$ is the same, and $B$ has the same form as for in the corresponding 3D box calculation. Plugging in $c_x = (3/4)(3/\pi)^{1/3}$, $c^{Dir}_{x,2} \approx 0.0767$, and $|\partial \Omega| = 6 |\Omega|^{2/3}$ for a cube, yields
\[
 B = \frac{3c^{Dir}_{x,2}|\partial \Omega|}{4c_x|\Omega|^{2/3}} = 6\Big(\frac \pi 3 \Big)^{1/3} c^{Dir}_{x,2}  \approx 0.47.
\]

\newpage
\section{Table S1: Energies of the particle in a box}\label{SI:Tabbox}

\begin{table}[ht]
{\setlength{\extrarowheight}{3pt}
    \begin{tabular}{c|c|c|c|c}
    N &$E(N)$& $\Tilde{E}(N)$  &$\Tilde{E}_{\rm d}(N)$& $\ENC(N) = \ETF(\NNC)$\\
    \hline
 1. & 4.93 & $1.64 \ (-66.67\%)$ & $4.11 \ (-16.67\%)$ & $5.55 \ (12.5\%)$ \\
 2. & 24.7 & $13.2 \ (-46.67\%)$ & $21.8 \ (-11.67\%)$ & $25.7 \ (4.17\%)$ \\
 3. & 69.1 & $44.4 \ (-35.71\%)$ & $62.9 \ (-8.93\%)$ & $70.5 \ (2.08\%)$ \\
 4. & 148 & $105 \ (-28.89\%)$ & $137 \ (-7.22\%)$ & $150 \ (1.25\%)$ \\
 5. & 271 & $206 \ (-24.24\%)$ & $255 \ (-6.06\%)$ & $274 \ (0.83\%)$ \\
 6. & 449 & $355 \ (-20.88\%)$ & $426 \ (-5.22\%)$ & $452 \ (0.6\%)$ \\
 7. & 691 & $564 \ (-18.33\%)$ & $659 \ (-4.58\%)$ & $694 \ (0.45\%)$ \\
 8. & 1007 & $842 \ (-16.34\%)$ & $966 \ (-4.08\%)$ & $1010 \ (0.35\%)$ \\
 9. & 1406 & $1199 \ (-14.74\%)$ & $1355 \ (-3.68\%)$ & $1410 \ (0.28\%)$ \\
 10. & 1900 & $1645 \ (-13.42\%)$ & $1836 \ (-3.35\%)$ & $1904 \ (0.23\%)$ \\
 11. & 2497 & $2189 \ (-12.32\%)$ & $2420 \ (-3.08\%)$ & $2502 \ (0.19\%)$ \\
 12. & 3208 & $2842 \ (-11.38\%)$ & $3116 \ (-2.85\%)$ & $3213 \ (0.16\%)$ \\
 13. & 4042 & $3614 \ (-10.58\%)$ & $3935 \ (-2.65\%)$ & $4047 \ (0.14\%)$ \\
 14. & 5009 & $4514 \ (-9.89\%)$ & $4885 \ (-2.47\%)$ & $5015 \ (0.12\%)$ \\
 15. & 6119 & $5552 \ (-9.27\%)$ & $5977 \ (-2.32\%)$ & $6126 \ (0.1\%)$ \\
 16. & 7382 & $6738 \ (-8.73\%)$ & $7221 \ (-2.18\%)$ & $7389 \ (0.09\%)$ \\
 17. & 8809 & $8082 \ (-8.25\%)$ & $8627 \ (-2.06\%)$ & $8816 \ (0.08\%)$ \\
 18. & 10407 & $9593 \ (-7.82\%)$ & $10204 \ (-1.96\%)$ & $10415 \ (0.07\%)$ \\
 19. & 12189 & $11283 \ (-7.44\%)$ & $11962 \ (-1.86\%)$ & $12197 \ (0.07\%)$ \\
 20. & 14163 & $13159 \ (-7.08\%)$ & $13912 \ (-1.77\%)$ & $14171 \ (0.06\%)$ \\
 21. & 16339 & $15234 \ (-6.77\%)$ & $16063 \ (-1.69\%)$ & $16348 \ (0.05\%)$ \\
 22. & 18728 & $17515 \ (-6.47\%)$ & $18424 \ (-1.62\%)$ & $18737 \ (0.05\%)$ \\
 23. & 21338 & $20014 \ (-6.21\%)$ & $21007 \ (-1.55\%)$ & $21348 \ (0.05\%)$ \\
 24. & 24181 & $22740 \ (-5.96\%)$ & $23820 \ (-1.49\%)$ & $24191 \ (0.04\%)$ \\
 25. & 27265 & $25702 \ (-5.73\%)$ & $26874 \ (-1.43\%)$ & $27275 \ (0.04\%)$ \\
 26. & 30601 & $28911 \ (-5.52\%)$ & $30178 \ (-1.38\%)$ & $30612 \ (0.04\%)$ \\
 27. & 34198 & $32377 \ (-5.32\%)$ & $33743 \ (-1.33\%)$ & $34209 \ (0.03\%)$ \\
 28. & 38067 & $36110 \ (-5.14\%)$ & $37578 \ (-1.29\%)$ & $38079 \ (0.03\%)$ \\
 29. & 42217 & $40118 \ (-4.97\%)$ & $41692 \ (-1.24\%)$ & $42229 \ (0.03\%)$ \\
 30. & 46659 & $44413 \ (-4.81\%)$ & $46097 \ (-1.2\%)$ & $46671 \ (0.03\%)$ \\
    \end{tabular}
}
\caption{Table that contains the data of Fig.~\ref{f:pboxintro}, with the first column being the number of electrons, the second column the exact energy, the third column the TF energy (red line), the fourth the TF energy obtained with the exact density (green line) and lastly the normalization-corrected TF (blue line) for the particle in a box. The percentage errors between the approximations and the exact can be found in between brackets}\label{t:partbox}
\end{table}
\newpage
\section{Table S2: Energies of the Bohr Atom}\label{SI:Tabbohr}

\begin{table}[ht]
{\setlength{\extrarowheight}{3pt}
    \begin{tabular}{c|c|c|c|c|c}
    N &$E(N)$& $\Tilde{E}(N)$ & $\Tilde{E}_{\rm sc}(N)$  &$\Tilde{E}_{\rm Z}(N)$& $\ENC(N) = \ETF(\NNC)$\\
    \hline
 1. & -0.500 & $-0.721 \ (44.2\%)$ & $-0.471 \ (-5.77\%)$ & $0.247 \ (-149\%)$& $-0.495 \ (-0.91\%)$ \\
 5. & -25.0 & $-30.8 \ (23.3\%)$ & $-24.6\ (-1.69\%)$ & $-22.6 \ (-9.77\%)$ & $-24.9 \ (-0.26\%)$\\
 14. & -294 & $-340 \ (15.9\%)$ & $-291 \ (-0.80\%)$ & $-282 \ (-4.02\%)$ & $-293 \ (-0.12\%)$\\
 30. & -1800 & $-2016 \ (12.0\%)$ & $-1791 \ (-0.46\%)$ & $-1760 \ (-2.19\%)$  & $-1798 \ (-0.069\%)$ \\
 55. & -7563 & $-8295 \ (9.70\%)$ & $-7539 \ (-0.30\%)$ & $-7457 \ (-1.39\%)$ & $-7559 \ (-0.045\%)$ \\
    \end{tabular}
}
\caption{Table that contains the data of Fig.~\ref{f:coulomb}, with the first column being the number of electrons, the second column the exact energy, the third column the TF energy (red line), the fourth Scott corrected TF energy, the fifth the TF energy with $Z$ fixed (blue line) and lastly the normalization-corrected TF (purple line) for the Bohr atom. The percentage errors between the approximations and the exact can be found in between brackets.}\label{t:Bohr}
\end{table}

\bibliography{Bibliography}

%% file: macros.tex
\def\bea{\begin{eqnarray}}
\def\eea{\end{eqnarray}}
\def\ben{\begin{equation}}
\def\bens{\begin{equation*}}
\def\een{\end{equation}}
\def\eens{\end{equation*}}
\def\benu{\begin{enumerate}}
\def\enu{\end{enumerate}}

\def\bei{\begin{itemize}}
\def\eei{\end{itemize}}
\def\beit{\begin{itemize}}
\def\eit{\end{itemize}}
\def\benu{\begin{enumerate}}
\def\enu{\end{enumerate}}

\def\sss{\scriptscriptstyle\rm}

\def\1var{(\bx_1...\bx\N)}

\def\bx{{\bf x}}

\def\x{_{\sss X}}

\def\N{_{\sss N}}

\def\LDA{^{\rm LDA}}

\def\sph_int{ {\int d^3 r}}

\newcounter{edit}

\newcommand{\kim}[1]{\textcolor{red}{#1}}
\newcommand{\cmmnt}[1]{}

\def\N{\mathcal{N}}



%% file: Bibliography.bib
@book{austen,
  author = {Jane Austen},
  year = {1813},
  title = {Pride and Prejudice},
  publisher = {T. Egerton, Whitehall}
}

@article{CorsoFriesecke,
    author = {Corso, Thiago Carvalho and Friesecke, Gero},
    title = {Next-order correction to the Dirac exchange energy of the free electron gas in the thermodynamic limit and generalized gradient approximations},
    journal = {Journal of Mathematical Physics},
    volume = {65},
    number = {8},
    pages = {081902},
    year = {2024},
    month = {08},
    issn = {0022-2488},
    doi = {10.1063/5.0152359},
    url = {https://doi.org/10.1063/5.0152359}


}

@article{Dir-MPCPS-30,
  doi = {10.1017/s0305004100016108},
  url = {https://doi.org/10.1017/s0305004100016108},
  year = {1930},
  month = jul,
  publisher = {Cambridge University Press ({CUP})},
  volume = {26},
  number = {3},
  pages = {376--385},
  author = {P. A. M. Dirac},
  title = {Note on Exchange Phenomena in the Thomas Atom},
  journal = {Mathematical Proceedings of the Cambridge Philosophical Society}
}

@article{Fer-MPCPS-27,
  year = {1927},
  volume = {6},
  pages = {602--607},
  author = {E. Fermi},
  title = {Un Metodo Statistico per la Determinazione di alcune Prioprietà dell’Atomo},
  journal = {Accademia Nazionale dei Lincei}
}

@article{Tho-MPCPS-27,
  doi = {10.1017/s0305004100011683},
  url = {https://doi.org/10.1017/s0305004100011683},
  year = {1927},
  month = jan,
  publisher = {Cambridge University Press ({CUP})},
  volume = {23},
  number = {5},
  pages = {542--548},
  author = {L. H. Thomas},
  title = {The calculation of atomic fields},
  journal = {Mathematical Proceedings of the Cambridge Philosophical Society}
}

@article{SimSonVucBur-JACS-22,
author = {Sim, Eunji and Song, Suhwan and Vuckovic, Stefan and Burke, Kieron},
title = {Improving Results by Improving Densities: Density-Corrected Density Functional Theory},
journal = {Journal of the American Chemical Society},
volume = {144},
number = {15},
pages = {6625-6639},
year = {2022},
doi = {10.1021/jacs.1c11506},
    note ={PMID: 35380807},

URL = { 
        https://doi.org/10.1021/jacs.1c11506
    
},
eprint = { 
        https://doi.org/10.1021/jacs.1c11506
    
}

}

@article{SonVucKimYuSimBur-Nat-23,
  doi = {10.1038/s41467-023-36094-y},
  url = {https://doi.org/10.1038/s41467-023-36094-y},
  year = {2023},
  month = feb,
  publisher = {Springer Science and Business Media {LLC}},
  volume = {14},
  number = {1},
  author = {Suhwan Song and Stefan Vuckovic and Youngsam Kim and Hayoung Yu and Eunji Sim and Kieron Burke},
  title = {Extending density functional theory with near chemical accuracy beyond pure water},
  journal = {Nature Communications}
}

@article{FraLar-Arx-24,
  title={Riesz means asymptotics for Dirichlet and Neumann Laplacians on Lipschitz domains},
  author={Frank, Rupert L and Larson, Simon},
  journal={Inventiones mathematicae},
  volume={241},
  number={3},
  pages={999--1079},
  year={2025},
  publisher={Springer}
}

@misc{SM,
      title={Supplemental Material: Approximate normalizations for approximate density functionals}, 
      author={{Adam Clay, Kiril Datchev, Wenlan Miao, Adam Wasserman, Kimberly J. Daas, and Kieron Burke}},
      year={2025},
      eprint={2504.03845},
      archivePrefix={arXiv},
      primaryClass={physics.chem-ph},
      url={https://arxiv.org/abs/2504.03845}, 
}

@article{GuiWan-JDG-12,
  title = {Semiclassical Spectral Invariants for Schr\"{o}dinger Operators},
  volume = {91},
  ISSN = {0022-040X},
  url = {http://dx.doi.org/10.4310/jdg/1343133702},
  DOI = {10.4310/jdg/1343133702},
  number = {1},
  journal = {Journal of Differential Geometry},
  publisher = {International Press of Boston},
  author = {Guillemin,  Victor and Wang,  Zuoqin},
  year = {2012},
  month = may 
}

@article{HelRob-AA-90,
  title = {Riesz means of bound states and semiclassical limit connected with a Lieb–Thirring’s conjecture},
  volume = {3},
  ISSN = {1875-8576},
  url = {http://dx.doi.org/10.3233/ASY-1990-3201},
  DOI = {10.3233/asy-1990-3201},
  number = {2},
  journal = {Asymptotic Analysis},
  publisher = {SAGE Publications},
  author = {Helffer,  B. and Robert,  D.},
  year = {1990},
  month = may,
  pages = {91–103}
}

@article{FraLar-AM-19,
  title = {Two-term spectral asymptotics for the Dirichlet Laplacian in a Lipschitz domain},
  volume = {2020},
  ISSN = {1435-5345},
  url = {http://dx.doi.org/10.1515/crelle-2019-0019},
  DOI = {10.1515/crelle-2019-0019},
  number = {766},
  journal = {Journal f\"{u}r die reine und angewandte Mathematik (Crelles Journal)},
  publisher = {Walter de Gruyter GmbH},
  author = {Frank,  Rupert L. and Larson,  Simon},
  year = {2019},
  month = aug,
  pages = {195–228}
}

@article{FraGei-BMS-12,
  title = {Semi-classical analysis of the Laplace operator with Robin boundary conditions},
  volume = {2},
  ISSN = {1664-3615},
  url = {http://dx.doi.org/10.1007/s13373-012-0028-5},
  DOI = {10.1007/s13373-012-0028-5},
  number = {2},
  journal = {Bulletin of Mathematical Sciences},
  publisher = {World Scientific Pub Co Pte Lt},
  author = {Frank,  Rupert L. and Geisinger,  Leander},
  year = {2012},
  month = oct,
  pages = {281–319}
}

@inproceedings{FraGei-MRQP-11,
  title = {TWO-TERM SPECTRAL ASYMPTOTICS FOR THE DIRICHLET LAPLACIAN ON A BOUNDED DOMAIN},
  url = {http://dx.doi.org/10.1142/9789814350365_0012},
  DOI = {10.1142/9789814350365_0012},
  booktitle = {Mathematical Results in Quantum Physics},
  publisher = {World Scientific},
  author = {Frank,  Rupert L. and Geisinger,  Leander},
  year = {2011},
  month = may 
}

@article{Ivr-FAA-80,
  title = {Second term of the spectral asymptotic expansion of the Laplace - Beltrami operator on manifolds with boundary},
  volume = {14},
  ISSN = {1573-8485},
  url = {http://dx.doi.org/10.1007/BF01086550},
  DOI = {10.1007/bf01086550},
  number = {2},
  journal = {Functional Analysis and Its Applications},
  publisher = {Springer Science and Business Media LLC},
  author = {Ivrii,  V. Ya.},
  year = {1980},
  pages = {98–106}
}

@article{DuiGui-IM-75,
  title = {The spectrum of positive elliptic operators and periodic bicharacteristics},
  volume = {29},
  ISSN = {1432-1297},
  url = {http://dx.doi.org/10.1007/BF01405172},
  DOI = {10.1007/bf01405172},
  number = {1},
  journal = {Inventiones Mathematicae},
  publisher = {Springer Science and Business Media LLC},
  author = {Duistermaat,  J. J. and Guillemin,  V. W.},
  year = {1975},
  month = feb,
  pages = {39–79}
}

@article{Wey-crll-13,
  title = {\"{U}ber die Randwertaufgabe der Strahlungstheorie und asymptotische Spektralgesetze.},
  volume = {1913},
  ISSN = {0075-4102},
  url = {http://dx.doi.org/10.1515/crll.1913.143.177},
  DOI = {10.1515/crll.1913.143.177},
  number = {143},
  journal = {crll},
  publisher = {Walter de Gruyter GmbH},
  author = {Weyl,  H.},
  year = {1913},
  pages = {177–202}
}

@article{HarProStu-IMRN-19,
  title = {Complementary Asymptotically Sharp Estimates for Eigenvalue Means of Laplacians},
  volume = {2021},
  ISSN = {1687-0247},
  url = {http://dx.doi.org/10.1093/imrn/rnz085},
  DOI = {10.1093/imrn/rnz085},
  number = {11},
  journal = {International Mathematics Research Notices},
  publisher = {Oxford University Press (OUP)},
  author = {Harrell II,  Evans M and Provenzano,  Luigi and Stubbe,  Joachim},
  year = {2019},
  month = may,
  pages = {8405–8450}
}

@article{PerRuzCsoVydScuConZhoBur-PRL-08,
  title = {Restoring the Density-Gradient Expansion for Exchange in Solids and Surfaces},
  author = {Perdew, John P. and Ruzsinszky, Adrienn and Csonka, G\'abor I. and Vydrov, Oleg A. and Scuseria, Gustavo E. and Constantin, Lucian A. and Zhou, Xiaolan and Burke, Kieron},
  journal = {Phys. Rev. Lett.},
  volume = {100},
  issue = {13},
  pages = {136406},
  numpages = {4},
  year = {2008},
  month = {Apr},
  publisher = {American Physical Society},
  doi = {10.1103/PhysRevLett.100.136406},
  url = {https://link.aps.org/doi/10.1103/PhysRevLett.100.136406}
}

@article{VucSonKozSimBur-JCTC-19,
  title={Density functional analysis: The theory of density-corrected DFT},
  author={Vuckovic, Stefan and Song, Suhwan and Kozlowski, John and Sim, Eunji and Burke, Kieron},
  journal={Journal of chemical theory and computation},
  volume={15},
  number={12},
  pages={6636--6646},
  year={2019},
  publisher={ACS Publications}
}

@BOOK{Esc-BOOK-96,
  title     = "The fundamentals of density functional theory",
  author    = "Eschrig, Helmut",
  publisher = "Vieweg+Teubner Verlag",
  series    = "Teubner Texte Zur Physik",
  month     =  jan,
  year      =  1996,
  address   = "Wiesbaden, Germany",
}

@article{Weyl-11,
author = {Weyl, H.},
journal = {Nachrichten von der Gesellschaft der Wissenschaften zu Göttingen, Mathematisch-Physikalische Klasse},
pages = {110-117},
title = {Ueber die asymptotische Verteilung der Eigenwerte},
url = {http://eudml.org/doc/58792},
volume = {1911},
year = {1911},
}

@BOOK{BenOrs-BOOK-99,
  title     = "Advanced mathematical methods for scientists and engineers {I}",
  author    = "Bender, Carl M and Orszag, Steven A",
  publisher = "Springer",
  month     =  oct,
  year      =  1999,
  address   = "New York, NY",
}

@BOOK{MasFed-BOOK-81,
  title     = "Semi-classical approximation in quantum mechanics",
  author    = "Maslov, Victor P and Fedoriuk, M V",
  publisher = "Kluwer Academic",
  series    = "Mathematical Physics and Applied Mathematics",
  edition   =  "1981",
  month     =  "aug",
  year      =  "1981",
  address   = "Dordrecht, Netherlands"
}

@article{Blo-ZfP-29,
  title = {Bemerkung zur Elektronentheorie des Ferromagnetismus und der elektrischen Leitfähigkeit},
  volume = {57},
  ISSN = {1434-601X},
  url = {http://dx.doi.org/10.1007/BF01340281},
  DOI = {10.1007/bf01340281},
  number = {7–8},
  journal = {Zeitschrift für Physik},
  publisher = {Springer Science and Business Media LLC},
  author = {Bloch,  F.},
  year = {1929},
  month = jul,
  pages = {545–555}
}

@article{RibLeeCanEllBur-PRL-15,
  title = {Corrections to Thomas-Fermi Densities at Turning Points and Beyond},
  volume = {114},
  ISSN = {1079-7114},
  url = {http://dx.doi.org/10.1103/PhysRevLett.114.050401},
  DOI = {10.1103/physrevlett.114.050401},
  number = {5},
  journal = {Physical Review Letters},
  publisher = {American Physical Society (APS)},
  author = {Ribeiro,  Raphael F. and Lee,  Donghyung and Cangi,  Attila and Elliott,  Peter and Burke,  Kieron},
  year = {2015},
  month = feb 
}

@article{EllPitGroBur-PRA-15,
  title = {Almost exact exchange at almost no computational cost in electronic structure},
  volume = {92},
  ISSN = {1094-1622},
  url = {http://dx.doi.org/10.1103/PhysRevA.92.022513},
  DOI = {10.1103/physreva.92.022513},
  number = {2},
  journal = {Physical Review A},
  publisher = {American Physical Society (APS)},
  author = {Elliott,  Peter and Cangi,  Attila and Pittalis,  Stefano and Gross,  E. K. U. and Burke,  Kieron},
  year = {2015},
  month = aug 
}

@article{OkuCanBur-PRB-24,
  title = {Orbital-free potential functionals with submillihartree errors for single-well slabs},
  volume = {109},
  ISSN = {2469-9969},
  url = {http://dx.doi.org/10.1103/PhysRevB.109.195156},
  DOI = {10.1103/physrevb.109.195156},
  number = {19},
  journal = {Physical Review B},
  publisher = {American Physical Society (APS)},
  author = {Okun,  Pavel and Cancio,  Antonio C. and Burke,  Kieron},
  year = {2024},
  month = may 
}

@article{EllLeeCanBur-PRL-08,
  title = {Semiclassical Origins of Density Functionals},
  author = {Elliott, Peter and Lee, Donghyung and Cangi, Attila and Burke, Kieron},
  journal = {Phys. Rev. Lett.},
  volume = {100},
  issue = {25},
  pages = {256406},
  numpages = {4},
  year = {2008},
  month = {Jun},
  publisher = {American Physical Society},
  doi = {10.1103/PhysRevLett.100.256406},
  url = {https://link.aps.org/doi/10.1103/PhysRevLett.100.256406}
}

@book{Eng-LNP-88,
    Author ={Berthold-Georg Englert},
title={Semiclassical Theory of Atoms},
  ISBN = {9783540391418},
  url = {http://dx.doi.org/10.1007/3-540-19204-2},
  DOI = {10.1007/3-540-19204-2},
  journal = {Lecture Notes in Physics},
  publisher = {Springer Berlin Heidelberg},
  year = {1988}
}

@article{FraMerSie-LMP-23,
  title = {The Scott conjecture for large Coulomb systems: a review},
  volume = {113},
  ISSN = {1573-0530},
  url = {http://dx.doi.org/10.1007/s11005-023-01631-9},
  DOI = {10.1007/s11005-023-01631-9},
  number = {1},
  journal = {Letters in Mathematical Physics},
  publisher = {Springer Science and Business Media LLC},
  author = {Frank,  Rupert L. and Merz,  Konstantin and Siedentop,  Heinz},
  year = {2023},
  month = jan 
}

@BOOK{CanCon-BOOK-99,
  title     = "Foliations, Volume 1",
  author    = "Candel, Alberto and Conlon, Lawrence",
  publisher = "American Mathematical Society",
  series    = "Graduate studies in mathematics",
  month     =  nov,
  year      =  1999,
  address   = "Providence, RI"
}

@incollection{OkuBur-arxiv-21,
  title = {SEMICLASSICS: THE HIDDEN THEORY BEHIND THE SUCCESS OF DFT},
  ISBN = {9789811272158},
  ISSN = {1793-0758},
  url = {http://dx.doi.org/10.1142/9789811272158_0007},
  DOI = {10.1142/9789811272158_0007},
  booktitle = {Density Functionals for Many-Particle Systems},
  publisher = {WORLD SCIENTIFIC},
  author = {Okun,  Pavel and Burke,  Kieron},
  year = {2023},
  month = feb,
  pages = {179–249}
}

@article{ColHin-JPCM-16,
  title = {Applications of large-scale density functional theory in biology},
  volume = {28},
  ISSN = {1361-648X},
  url = {http://dx.doi.org/10.1088/0953-8984/28/39/393001},
  DOI = {10.1088/0953-8984/28/39/393001},
  number = {39},
  journal = {Journal of Physics: Condensed Matter},
  publisher = {IOP Publishing},
  author = {Cole,  Daniel J and Hine,  Nicholas D M},
  year = {2016},
  month = aug,
  pages = {393001}
}

@article{Lie-RMP-81,
  title = {Thomas-fermi and related theories of atoms and molecules},
  author = {Lieb, Elliott H.},
  journal = {Rev. Mod. Phys.},
  volume = {53},
  issue = {4},
  pages = {603--641},
  numpages = {0},
  year = {1981},
  month = {Oct},
  publisher = {American Physical Society},
  doi = {10.1103/RevModPhys.53.603},
  url = {https://link.aps.org/doi/10.1103/RevModPhys.53.603}
}

@article{CanCheKruBur-JCP-18,
author = {Cancio,Antonio  and Chen,Guo P.  and Krull,Brandon T.  and Burke,Kieron },
title = {Fitting a round peg into a round hole: Asymptotically correcting the generalized gradient approximation for correlation},
journal = {The Journal of Chemical Physics},
volume = {149},
number = {8},
pages = {084116},
year = {2018},
doi = {10.1063/1.5021597},

URL = { 
        https://doi.org/10.1063/1.5021597
    
},
eprint = { 
        https://doi.org/10.1063/1.5021597
    
}

}

@article{AdeChiAdeSadHamRay-Pha-22,
  title = {Application of DFT Calculations in Designing Polymer-Based Drug Delivery Systems: An Overview},
  volume = {14},
  ISSN = {1999-4923},
  url = {http://dx.doi.org/10.3390/pharmaceutics14091972},
  DOI = {10.3390/pharmaceutics14091972},
  number = {9},
  journal = {Pharmaceutics},
  publisher = {MDPI AG},
  author = {Adekoya,  Oluwasegun Chijioke and Adekoya,  Gbolahan Joseph and Sadiku,  Emmanuel Rotimi and Hamam,  Yskandar and Ray,  Suprakas Sinha},
  year = {2022},
  month = sep,
  pages = {1972}
}

@article{DawDegSteNajRatGen-WIR-22,
author = {Dawson, William and Degomme, Augustin and Stella, Martina and Nakajima, Takahito and Ratcliff, Laura E. and Genovese, Luigi},
title = {Density functional theory calculations of large systems: Interplay between fragments, observables, and computational complexity},
journal = {WIREs Computational Molecular Science},
volume = {12},
number = {3},
pages = {e1574},
doi = {https://doi.org/10.1002/wcms.1574},
year = {2022}
}

@article{KimSimBur-JCP-14,
  title={Ions in solution: Density corrected density functional theory (DC-DFT)},
  author={Kim, Min-Cheol and Sim, Eunji and Burke, Kieron},
  journal={J. Chem. Phys.},
  volume={140},
  number={18},
  pages={18A528},
  year={2014},
  publisher={AIP}
}

@article{SonVucSimBur-JCTC-22,
author = {Song, Suhwan and Vuckovic, Stefan and Sim, Eunji and Burke, Kieron},
title = {Density-Corrected DFT Explained: Questions and Answers},
journal = {Journal of Chemical Theory and Computation},
volume = {18},
number = {2},
pages = {817-827},
year = {2022},
doi = {10.1021/acs.jctc.1c01045},
    note ={PMID: 35048707},

URL = { 
        https://doi.org/10.1021/acs.jctc.1c01045
    
},
eprint = { 
        https://doi.org/10.1021/acs.jctc.1c01045
    
}

}

@article{HorDwaPer-NCS-21,
  title = {Promises and perils of computational materials databases},
  volume = {1},
  ISSN = {2662-8457},
  url = {http://dx.doi.org/10.1038/s43588-020-00016-5},
  DOI = {10.1038/s43588-020-00016-5},
  number = {1},
  journal = {Nature Computational Science},
  publisher = {Springer Science and Business Media LLC},
  author = {Horton,  M. K. and Dwaraknath,  S. and Persson,  K. A.},
  year = {2021},
  month = jan,
  pages = {3–5}
}

@article{HeiLie-PRA-95,
  title = {Electron density near the nucleus of a large atom},
  author = {Heilmann, Ole J. and Lieb, Elliott H.},
  journal = {Phys. Rev. A},
  volume = {52},
  issue = {5},
  pages = {3628--3643},
  numpages = {0},
  year = {1995},
  month = {Nov},
  publisher = {American Physical Society},
  doi = {10.1103/PhysRevA.52.3628},
  url = {https://link.aps.org/doi/10.1103/PhysRevA.52.3628}
}

@article{HohKoh-PR-64,
	Author = {Hohenberg, P. and Kohn, W.},
	Journal = {Phys. Rev.},
	Pages = {B 864},
	Title = {Inhomogeneous Electron Gas},
	Volume = {{136}},
	Year = {1964}}

@article{Sco-PMJC-52,
  title = {LXXXII. The binding energy of the Thomas-Fermi Atom},
  volume = {43},
  ISSN = {1941-5990},
  url = {http://dx.doi.org/10.1080/14786440808520234},
  DOI = {10.1080/14786440808520234},
  number = {343},
  journal = {The London,  Edinburgh,  and Dublin Philosophical Magazine and Journal of Science},
  publisher = {Informa UK Limited},
  author = {Scott,  J.M.C.},
  year = {1952},
  month = aug,
  pages = {859–867}
}

@article{ArgRedCanBur-PRL-22,
  title = {Leading Correction to the Local Density Approximation for Exchange in Large-$Z$ Atoms},
  author = {Argaman, Nathan and Redd, Jeremy and Cancio, Antonio C. and Burke, Kieron},
  journal = {Phys. Rev. Lett.},
  volume = {129},
  issue = {15},
  pages = {153001},
  numpages = {6},
  year = {2022},
  month = {Oct},
  publisher = {American Physical Society},
  doi = {10.1103/PhysRevLett.129.153001},
  url = {https://link.aps.org/doi/10.1103/PhysRevLett.129.153001}
}

@article{KimSimBur-PRL-13,
	Author = {Kim, Min-Cheol and Sim, Eunji and Burke, Kieron},
    Title = {Understanding and Reducing Errors in Density Functional Calculations},
	Issue = {7},
	Journal = {Phys. Rev. Lett.},
	Pages = {073003},
	Volume = {111},
	Year = {2013}}

@article{MarPla-MPS-56,
    Author= {N. H. March and John Stanley Plaskett},
  volume = {235},
  ISSN = {2053-9169},
    title={The relation between the Wentzel-Kramers-Brillouin and the Thomas-Fermi approximations},
  url = {http://dx.doi.org/10.1098/rspa.1956.0094},
  DOI = {10.1098/rspa.1956.0094},
  number = {1202},
  journal = {Proceedings of the Royal Society of London. Series A. Mathematical and Physical Sciences},
  publisher = {The Royal Society},
  year = {1956},
  month = may,
  pages = {419–431}
}

@article{ConFabLarDel-PRL-11,
  title = {Semiclassical Neutral Atom as a Reference System in Density Functional Theory},
  volume = {106},
  ISSN = {1079-7114},
  url = {http://dx.doi.org/10.1103/PhysRevLett.106.186406},
  DOI = {10.1103/physrevlett.106.186406},
  number = {18},
  journal = {Physical Review Letters},
  publisher = {American Physical Society (APS)},
  author = {Constantin,  Lucian A. and Fabiano,  E. and Laricchia,  S. and Della Sala,  F.},
  year = {2011},
  month = may 
}

@article{MiLuoTriPav-CR-23,
  title = {Orbital-Free Density Functional Theory: An Attractive Electronic Structure Method for Large-Scale First-Principles Simulations},
  volume = {123},
  ISSN = {1520-6890},
  url = {http://dx.doi.org/10.1021/acs.chemrev.2c00758},
  DOI = {10.1021/acs.chemrev.2c00758},
  number = {21},
  journal = {Chemical Reviews},
  publisher = {American Chemical Society (ACS)},
  author = {Mi,  Wenhui and Luo,  Kai and Trickey,  S. B. and Pavanello,  Michele},
  year = {2023},
  month = oct,
  pages = {12039–12104}
}

@article{BerBur-JPA-20,
  title = {Exact and approximate energy sums in potential wells},
  volume = {53},
  ISSN = {1751-8121},
  url = {http://dx.doi.org/10.1088/1751-8121/ab69a6},
  DOI = {10.1088/1751-8121/ab69a6},
  number = {9},
  journal = {Journal of Physics A: Mathematical and Theoretical},
  publisher = {IOP Publishing},
  author = {Berry,  M V and Burke,  Kieron},
  year = {2020},
  month = feb,
  pages = {095203}
}

@article{KohSha-PR-65,
	Author = {Kohn, W. and Sham, L. J.},
	Journal = {Phys. Rev.},
	Pages = {A 1133},
	Title = {Self-Consistent Equations Including Exchange and Correlation Effects},
	Volume = {140},
	Year = {1965}}

@article{Sla-PR-51,
	Author = {Slater, John C},
	Journal = {Phys. Rev.},
	Number = {3},
	Pages = {385},
	Publisher = {APS},
	Title = {A simplification of the Hartree-Fock method},
	Volume = {81},
	Year = {1951}}

@article{DelSchWenMesHutVan-CPC-15,
title = {Enabling simulation at the fifth rung of DFT: Large scale RPA calculations with excellent time to solution},
journal = {Computer Physics Communications},
volume = {187},
pages = {120-129},
year = {2015},
issn = {0010-4655},
doi = {https://doi.org/10.1016/j.cpc.2014.10.021},
url = {https://www.sciencedirect.com/science/article/pii/S0010465514003671},
author = {Mauro {Del Ben} and Ole Schütt and Tim Wentz and Peter Messmer and Jürg Hutter and Joost VandeVondele}
}

@article{SunRuzPer-PRL-15,
	Author = {Sun, Jianwei and Ruzsinszky, Adrienn and Perdew, John P},
	Journal = {Phys. Rev. Lett.},
	Number = {3},
	Pages = {036402},
	Publisher = {APS},
	Title = {Strongly constrained and appropriately normed semilocal density functional},
	Volume = {115},
	Year = {2015}}

@book{Fri-GTP-17,
  title = {Theoretical Atomic Physics},
  ISBN = {9783319477695},
  ISSN = {1868-4521},
  url = {http://dx.doi.org/10.1007/978-3-319-47769-5},
  DOI = {10.1007/978-3-319-47769-5},
  journal = {Graduate Texts in Physics},
  publisher = {Springer International Publishing},
  author = {Friedrich,  Harald},
  year = {2017}
}

@article{Sch-PRA-81,
  title = {Thomas-Fermi model: The second correction},
  volume = {24},
  ISSN = {0556-2791},
  url = {http://dx.doi.org/10.1103/PhysRevA.24.2353},
  DOI = {10.1103/physreva.24.2353},
  number = {5},
  journal = {Physical Review A},
  publisher = {American Physical Society (APS)},
  author = {Schwinger,  Julian},
  year = {1981},
  month = nov,
  pages = {2353–2361}
}

@article{Sch-PRA-80,
  title = {Thomas-Fermi model: The leading correction},
  volume = {22},
  ISSN = {0556-2791},
  url = {http://dx.doi.org/10.1103/PhysRevA.22.1827},
  DOI = {10.1103/physreva.22.1827},
  number = {5},
  journal = {Physical Review A},
  publisher = {American Physical Society (APS)},
  author = {Schwinger,  Julian},
  year = {1980},
  month = nov,
  pages = {1827–1832}
}

@book{EngDre-BOOK-11,
author="Eberhard Engel, Reiner M. Dreizler",
title="Density Functional Theory: An Advanced Course",
bookTitle="Theoretical, Mathematical \& Computational Physics",
year="2011",
publisher="Springer-Verlag Berlin Heidelberg",
address="Berlin, Heidelberg",
isbn="978-3-642-14090-7",
doi="10.1007/978-3-642-14090-7",
url="http://dx.doi.org/10.1007/978-3-642-14090-7"
}

@article{WasPav-IJQC-20,
author = {Wasserman, Adam and Pavanello, Michele},
title = {Quantum embedding electronic structure methods},
journal = {International Journal of Quantum Chemistry},
volume = {120},
number = {21},
pages = {e26495},
doi = {https://doi.org/10.1002/qua.26495},
url = {https://onlinelibrary.wiley.com/doi/abs/10.1002/qua.26495},
eprint = {https://onlinelibrary.wiley.com/doi/pdf/10.1002/qua.26495},
year = {2020}
}

@misc{AreNitPetSte-MAEIC-09,
  title = {Weyl’s Law: Spectral Properties of the Laplacian in Mathematics and Physics},
  ISBN = {9783527628025},
  url = {http://dx.doi.org/10.1002/9783527628025.ch1},
  DOI = {10.1002/9783527628025.ch1},
  journal = {Mathematical Analysis of Evolution,  Information,  and Complexity},
  publisher = {Wiley},
  author = {Arendt,  Wolfgang and Nittka,  Robin and Peter,  Wolfgang and Steiner,  Frank},
  year = {2009},
  month = mar,
  pages = {1–71}
}

@article{LieSim-PRL-73,
  title={Thomas-Fermi theory revisited},
  author={Lieb, Elliott H and Simon, Barry},
  journal={Physical Review Letters},
  volume={31},
  number={11},
  pages={681},
  year={1973},
  publisher={APS}
}

@article{Tit-QJM-54,
  title = {ON THE ASYMPTOTIC DISTRIBUTION OF EIGENVALUES},
  volume = {5},
  ISSN = {1464-3847},
  url = {http://dx.doi.org/10.1093/qmath/5.1.228},
  DOI = {10.1093/qmath/5.1.228},
  number = {1},
  journal = {The Quarterly Journal of Mathematics},
  publisher = {Oxford University Press (OUP)},
  author = {Titchmarsh,  E. C.},
  year = {1954},
  pages = {228–240}
}

@article{CanLeeEllBur-PRB-10,
  title = {Leading corrections to local approximations},
  author = {Cangi, Attila and Lee, Donghyung and Elliott, Peter and Burke, Kieron},
  journal = {Phys. Rev. B},
  volume = {81},
  issue = {23},
  pages = {235128},
  numpages = {14},
  year = {2010},
  month = {Jun},
  publisher = {American Physical Society},
  doi = {10.1103/PhysRevB.81.235128},
  url = {https://link.aps.org/doi/10.1103/PhysRevB.81.235128}
}

@article{BurCanGouPit-JCP-16,
Pub-num={173},
  author = "Burke, Kieron and Cancio, Antonio and Gould, Tim and Pittalis, Stefano",
   title = "Locality of correlation in density functional theory",
   journal = "The Journal of Chemical Physics",
   year = "2016",
   volume = "145",
   number = "5", 
   pages = "054112",
   url = "http://scitation.aip.org/content/aip/journal/jcp/145/5/10.1063/1.4959126",
   doi = "http://dx.doi.org/10.1063/1.4959126" 
}

@article{Bur-FD-20,
  title = {Deriving approximate functionals with asymptotics},
  volume = {224},
  ISSN = {1364-5498},
  url = {http://dx.doi.org/10.1039/d0fd00057d},
  DOI = {10.1039/d0fd00057d},
  journal = {Faraday Discussions},
  publisher = {Royal Society of Chemistry (RSC)},
  author = {Burke,  Kieron},
  year = {2020},
  pages = {98–125}
}

@book{watson,
  author = {G. N. Watson},
  year = {1944},
  title = {A Treatise on the Theory of Bessel Functions},
  publisher = {Cambridge University Press}
}

@book{dimassisjostrand,
  author = {M. Dimassi and J. Sj\"ostrand},
  year = {1999},
  title = {Spectral Asymptotics in the Semi-Classical Limit},
  publisher = {Cambridge University Press}
}

@book{zworski,
  author = {Maciej Zworski},
  year = {2012},
  title = {Semiclassical Analysis},
  publisher = {American Mathematical Society}
}

@book{guilleminsternberg,
  author = {Victor Guillemin and Sholomo Sternberg},
  year = {2013},
  title = {Semi-Classical Analysis},
  publisher = {International Press}
}

@article{ivrii100,
  title = {100 years of Weyl’s law},
  volume = {6},
  ISSN = {1664-3615},
  url = {https://doi.org/10.1007/s13373-016-0089-y},
  DOI = {10.1007/s13373-016-0089-y},
  number = {3},
  journal = {Bulletin of Mathematical Sciences},
  publisher = {Springer},
  author = {Victor Ivrii},
  year = {2016},
  pages = {379--452}
}
